\documentclass[twocolumn,preprintnumbers,amsmath,amssymb,superscriptaddress,longbibliography,nofootinbib]{revtex4-1}
\usepackage{filecontents}
\usepackage{graphicx}
\usepackage{bm}
\usepackage{dsfont}
\usepackage[usenames,dvipsnames]{xcolor}
\usepackage{pstricks}
\usepackage[tight]{subfigure}
\usepackage{verbatim}
\usepackage{units}
\usepackage{multirow}
\usepackage{enumitem}
\usepackage{mathrsfs}
\usepackage{leftidx}
\usepackage{xspace}
\usepackage{braket}
\usepackage{bbm}
\usepackage[normalem]{ulem}
\usepackage{cancel}
\usepackage{amsfonts,amssymb,amsmath}

\usepackage{xr}

\usepackage[a4paper]{hyperref}
\hypersetup{colorlinks=true,linktoc=all,linkcolor=blue,breaklinks=true,citecolor=blue,urlcolor=blue}

\usepackage[T1]{fontenc}
\usepackage[utf8]{inputenc}
\usepackage[russian,english]{babel}

\usepackage[babel,english=american]{csquotes}

\usepackage{cleveref}
\newcommand{\T}{\mathcal{T}}

\newcommand{\kB}{k_\text{B}}

\newcommand{\tr}[1]{\text{tr}\left\{ {#1} \right\}}

\newcommand{\be}{\begin{equation}\small\begin{aligned}}{}
\newcommand{\ee}{\end{aligned}\end{equation}}

\newcommand{\MRL}{\mathrm{L}}

\newcommand{\MRS}{\mathrm{S}}
\newcommand{\MRR}{\mathrm{R}}
\newcommand{\MBK}{\mathbf{K}}
\newcommand{\MBV}{\mathbf{V}}

\newcommand{\AGKSL}{\mathcal{A}^\mathrm{GKSL}}
\newcommand{\KGKSL}{\mathcal{K}^\mathrm{GKSL}}

\begin{document}
\title{Impact of Nonlinearities on Local Kinetic and Thermokinetic Uncertainty Relations in Bosonic Transport}

\author{Didrik Palmqvist}
\affiliation{Department of Microtechnology and Nanoscience (MC2), Chalmers University of Technology, S-412 96 G\"oteborg, Sweden}

\author{Luca Magazz\`u}
\affiliation{Pico group, QTF Centre of Excellence, Department of Applied Physics,
Aalto University School of Science, P.O. Box 13500, 00076 Aalto, Finland}

\author{Milena Grifoni}
\affiliation{Institute for Theoretical Physics, University of Regensburg, 93040 Regensburg, Germany}
\affiliation{Halle-Berlin-Regensburg Cluster of Excellence CCE,
University of Regensburg, 93040 Regensburg, Germany}

\author{Janine Splettstoesser}
\affiliation{Department of Microtechnology and Nanoscience (MC2),Chalmers University of Technology, S-412 96 G\"oteborg, Sweden}

\date{\today}

\begin{abstract}
The precision of transport observables can be bounded by the entropy production and the activity of the transport process via kinetic and thermokinetic uncertainty relations. In linear bosonic systems, such uncertainty relations can provide tight bounds when the activity is replaced by a \textit{local} activity of the measurement contact of interest---even in the strong-coupling regime. How much nonlinearities (or interactions) impact the validity and predictiveness of these local bounds and how much this impact depends on the concrete definition of activity are open questions that we address for two experimentally relevant model systems, a harmonic oscillator network and the intrinsically nonlinear spin-boson model. We thereby provide experimentally testable predictions on how nonlinearities impact or even break local kinetic and thermokinetic precision bounds. 
\end{abstract}

\maketitle

\section{Introduction}

\begin{figure}[b!]
 \centering
 \includegraphics[width=2.5in]{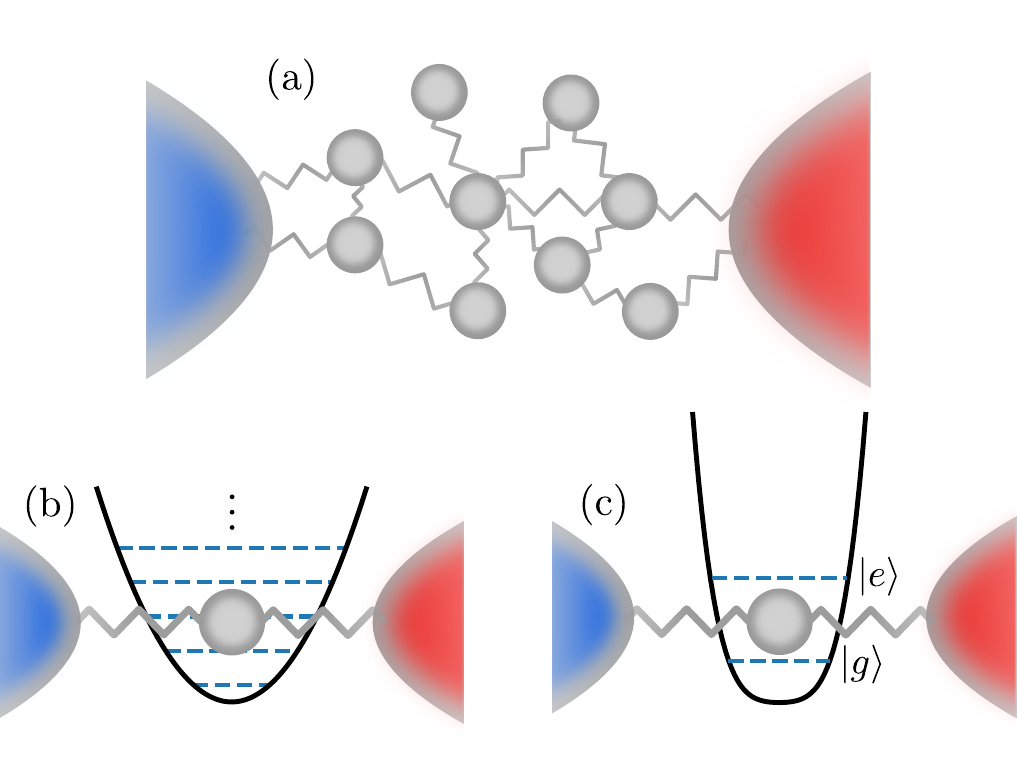}
\caption{(a) Sketch of a network of harmonic oscillators coupled to two bosonic reservoirs (or, equivalently, baths), where colors indicate possibly different temperatures. (b)~Limit of a single harmonic oscillator connected to two reservoirs. (c)~In the presence of strong nonlinearities, the central region is modeled by a two-level system---the nonequilibrium spin-boson model (NESB).}
 \label{fig: system sketch}
\end{figure}

Processes in small-scale quantum devices typically display fluctuations that can be significant compared to the desired average of an observable~\cite{Blanter2000Sep,Esposito2009Dec}. It is therefore crucial to understand which effects or physical properties of a system constrain the {\em precision} that a process can possibly reach. Recently developed thermodynamic~\cite{Barato2015Apr,Gingrich2016Mar, Horowitz2020Jan,Koyuk2020Dec}, kinetic~\cite{DiTerlizzi2018Dec,Garrahan2017Mar}, and thermokinetic~\cite{Vo2022Sep} uncertainty relations provide such constraints in terms of the entropy production or of the activity of a process. While these bounds were initially developed for classical stochastic processes, their validity for quantum systems and possible extensions to the quantum regime have been studied extensively during the last few years using weak-coupling Markovian master equations~\cite{Hasegawa2020Jul,Hasegawa2023May,Prech2025Jan,VanVu2025Mar}, strongly coupled conductors but with non-interacting or weakly interacting particles using scattering theory~\cite{brandner_thermodynamic_2018,potanina_thermodynamic_2021,Palmqvist2025Oct,Brandner2025Feb,Palmqvist2025Jul,Blasi2026Jul,Tesser2026Jul}, frameworks based on repeated measurements~\cite{Hasegawa2025Nov,VanVu2025Aug}, or on information-geometric approaches~\cite{Hasegawa2023May,Nishiyama2024Apr,Palmqvist2026Jul}.

In order to extend \textit{kinetic} or \textit{thermokinetic} uncertainty relations to the quantum regime, different types of activity definitions or activity estimates have been proposed~\cite{Hasegawa2023May,Nishiyama2024Apr,Blasi2026Jul,Palmqvist2026,Palmqvist2026Jul}. 
In fermionic coherent-transport systems, quantum effects can lead to a violation of the classical bounds through suppressed noise.
By contrast, super-Poissonian noise, often occurring in bosonic systems, rather deteriorates precision and tends to loosen the precision constraints given by uncertainty relations~\cite{Saryal2019Oct,Palmqvist2025Oct,Palmqvist2025Jul,Sobrino2026Jul}. However, it has recently been shown that the thermokinetic cost for precision can be decreased in bosonic systems in the presence of driving inducing quantum-coherent oscillations between energy eigenstates~\cite{Prech2025Jan}. Also, indications for improved relations between precision and its cost have been found for a \textit{nonlinear} system coupled to a single bosonic bath~\cite{Prech2025Jan}. This is particularly relevant since nonlinear bosonic systems play a crucial role in future quantum technologies~\cite{Kurizki2015Mar,Devoret2013Mar,Blais2007Mar}. 

In steady-state transport, it has furthermore been found that replacing the activity in kinetic or thermokinetic uncertainty relations by a \textit{local} activity improves how predictive---or how tight---precision bounds are~\cite{Palmqvist2025Oct,Palmqvist2025Jul,Blasi2026Jul,Palmqvist2026Jul}. In the context of thermodynamics, steady-state transport is, for example, of high interest in small-scale thermoelectric heat engines~\cite{Josefsson2018Oct} or in qubit-based heat valves~\cite{Ronzani2018,Senior2020}. Therefore, an important question is, how nonlinearities in bosonic systems---for example, due to interactions or Kerr nonlinearities---impact \textit{local} precision constraints.

In this paper, we show how local precision bounds of the type of kinetic or thermokinetic uncertainty relations derived for linear systems~\cite{Palmqvist2025Oct,Palmqvist2025Jul} are broken or how their predictiveness is impacted~\cite{Palmqvist2026Jul} in steady-state bosonic transport, when the system is characterized by nonlinearities---here exemplified with the precision of energy transport in the nonequilibrium spin-boson model (NESB), see Fig.~\ref{fig: system sketch}. As a \textit{linear} reference system, we consider uncertainty relations in a network of harmonic oscillators (panels (a) and (b) of Fig.~\ref{fig: system sketch}). We compare the predictiveness of different activity or activity-like definitions, which have until now mostly been employed for fermionic systems. We then demonstrate how the NESB model impacts the local bounds for any of the activity definitions or even breaks those whose validity is guaranteed for linear systems only.

Energy transport in the nonequilibrium spin-boson model is experimentally realized in superconducting qubit systems coupled to cavities or waveguides~\cite{Ronzani2018, Senior2020,Pekola2021}. We hence expect that the breaking of local uncertainty relations predicted here can be experimentally tested.

The paper is organized as follows. We introduce the precision bounds of interest in Sec.~\ref{sec:intro_tradeoffs} and provide details about the recently discussed activity measures for steady-state transport, which will be used in this paper, in Sec.~\ref{sec:dynamical_activity}. In Sec.~\ref{sec:HO}, we present the model for the harmonic-oscillator network together with a full-counting-statistics calculation of the energy transport in this system. We then analyze how its precision performs with respect to the precision bounds, where we compare the predictiveness of different activity definitions for broad parameter regimes. The main results about the breaking of local kinetic and thermokinetic precision bounds in the spin-boson model are discussed in Sec.~\ref{sec:NESB} using a Gorini–Kossakowski–Sudarshan–Lindblad (GKSL) master equation approach.

\section{Trade-off relations and Dynamical activity}

\subsection{Trade-off relations}\label{sec:intro_tradeoffs}

Recently, a number of trade-off relations bounding the achievable precision of a process have been put forward, initially for classical stochastic processes, but strong efforts are currently being put into extending them even to quantum systems~\cite{potanina_thermodynamic_2021,Brandner2025Feb,Brandner2025Oct,Prech2025Jan,Blasi2026Jul,Hasegawa2020Jul,Hasegawa2023May,Hasegawa2025Nov,VanVu2025Mar,VanVu2025Aug,Palmqvist2026Jul}. In this section, we review the steady-state versions of some of these bounds for the precision $\mathcal{P}$. For an arbitrary current $I^{(X)}$ transporting some quantity $X$ with current fluctuations $S^{(X)}$, see Eq.~\eqref{eq:SN} below, the precision\footnote{Note that the precision has the dimension of an inverse time.} is defined as $\mathcal{P}^{(X)}:= (I^{(X)})^2/S^{(X)}$.

The \textit{thermodynamic} uncertainty relation (TUR) bounds the precision by the total entropy production rate $\sigma$, 
\begin{equation}\label{eq:TUR_intro}
 \mathcal{P}^{(X)} \leq \frac{\sigma}{2\kB}\,,
\end{equation}
with $\kB$ the Boltzmann constant\footnote{The standard TUR requries that $I^{(X)}$ is a thermodynamic current with antisymmetric jump weights.}.
 In steady-state transport settings, the total entropy production rate can be shown to be given by the entropy production rate in the baths alone~\cite{Esposito2010Jan,Tesser2026Apr}. For large thermal baths (thermodynamic limit), this is obtained from the Clausius relation connecting entropy production to heat exchange and bath temperatures. Here we consider bosonic reservoirs where the excitation number is not conserved, e.g. photonic or phononic, and therefore the chemical potential equals zero, $\mu=0$, meaning that the energy and heat currents are the same. Then, the total entropy-production rate reads
\begin{equation}
 \sigma = -\sum_\alpha \frac{I^{(E)}_\alpha}{T_\alpha}\,,\label{eq:entropy_clausius}
\end{equation}
with energy current $I^{(E)}_\alpha$ and temperature $T_\alpha$ in bath $\alpha$.

While TURs are valid in a broad range of nonequilibrium settings, they are usually closer to saturation around equilibrium. A different constraint, which is often more predictive, i.e. tighter, far from equilibrium, is the kinetic uncertainty relation (KUR) 
\begin{equation}\label{eq:KUR_intro}
 \mathcal{P}^{(X)} \leq \mathcal{K},
\end{equation}
bounding precision by the activity rate $\mathcal{K}$. For classical, continuous-time Markov processes where the KUR was initially discovered~\cite{Garrahan2017Mar,DiTerlizzi2018Dec}, the dynamical activity counts the average number of transitions (or jumps) that a system has undergone during its evolution~\cite{Garrahan2007May,Lecomte2007Apr,Maes2020Mar}. Extending this concept to quantum dynamics is, however, in general challenging. We will present different routes of relevance for quantum transport in the following section, Sec.~\ref{sec:dynamical_activity}. 

The crossover between the KUR and TUR is found in a unified thermokinetic uncertainty relation, taking the form~\cite{Vo2022Sep,VanVu2025Mar,Palmqvist2026}
\begin{equation}\label{eq:TKUR_intro}
 \mathcal{P}^{(X)} \leq \frac{\sigma}{2\kB} \Xi\left[\frac{\sigma}{2\kB \mathcal{K}}\right].
\end{equation}
Here $\Xi[x] := x/\Omega^2[x]$ occurs, where $\Omega[x]$ is the inverse function of $x \tanh[x]$. In the following, these bounds will be analyzed for different bosonic transport models, both with and without nonlinearities.

In linear bosonic systems, two important refinements of the above bounds have been found. First, a stricter constraint on entropy production compared to the TUR has been established~\cite{Palmqvist2025Jul} 
\begin{equation}\label{eq:bosonic linear bound}
 \begin{aligned}
 &\mathcal{P}^{(X)} \leq C(S^{(N)},\mathcal{P}^{(X)} )\leq \frac{\sigma}{2\kB},\\
 &C(y,z):= \frac{1}{2}\ln\left[\frac{\sqrt{y} + \sqrt{z}}{\sqrt{y} - \sqrt{z}} \right] \sqrt{yz}.
 \end{aligned}
\end{equation}
Here, the particle current noise is used to estimate the rates of single-particle transfers between reservoirs. 
Secondly, it was shown that the kinetic and thermokinetic bounds become tighter when the activities $\mathcal{K}$ are replaced by \textit{local} activities $\mathcal{K}_\alpha$, which only count quasiparticle exchange with the reservoir in which the current of interest is measured. That this is possible is, in some approaches~\cite{Palmqvist2025Jul,Palmqvist2025Oct,Blasi2026Jul}, due to the linear nature of the system (discussed below for the example of a harmonic network); if one instead treats interacting or nonlinear bosonic transport, the local bounds might be violated. By contrast, the local susceptibility-KUR~\cite{Palmqvist2026Jul} continues to hold even for nonlinear Hamiltonians; however, the impact of nonlinearities on its predictiveness has not been analyzed until now. We also point out that for noninteracting \textit{fermionic} quantum transport, both the KUR using the local and global activities are violated in general~\cite{Prech2025Jan}.

\subsection{Dynamical activity}\label{sec:dynamical_activity}

The dynamical activity introduced in Sec.~\ref{sec:intro_tradeoffs} above is a key quantity occurring in KURs and TKURs. As the name suggests, it is intuitively related to how active a system is. In \textit{classical} stochastic thermodynamics, it is measured in terms of jumps between states. The generalisation of activity to quantum systems is an ongoing research question, and there exist multiple suggestions in the literature~\cite{Hasegawa2023May, Hasegawa2021Jan, VanVu2025Mar, Tesser2025, Blasi2026Jul, Palmqvist2026,Palmqvist2026Jul}.
In this section, we give an overview, derivations, and motivations for activity-like measures that we will use as the basis of the study presented in this paper.

\subsubsection{Dynamical activity from a (quantum) jump approach}

We start by introducing activity for an open quantum system interacting weakly with Markovian baths, where stochastic jumps are well defined. Such a system is described by a GKSL master equation for the reduced density operator
\begin{equation}\label{eq:GKSL}
 \dot {\hat{\rho}}_\MRS(t) =\mathcal{L} \hat{\rho}_\MRS(t)= -\frac{i}{\hbar} [\hat{H}_\MRS(t),\hat{\rho}_\MRS(t)] + \sum_{\alpha, j} \mathcal{D}[\hat{L}_{\alpha, j}(t)] \hat{\rho}_\MRS(t).
\end{equation}
Here, $\mathcal{L}$ is the Liouvillian superoperator, $\hat{H}_\MRS(t)$ is the system Hamiltonian, and $\hat{L}_{\alpha, j}$ is a jump operator associated with bath $\alpha$. The dissipators are defined as $\mathcal{D}[\hat{L}]\hat{\rho} := \hat{L} \hat{\rho} \hat{L}^\dag - \frac{1}{2} \{\hat{L}^\dag \hat{L},\hat{\rho}\}$, where the curly brackets denote the anti-commutator. For a jump unraveling of this master equation, the jump operators are used to generalise the notion of a counting variable $X$ with average~\cite{Landi2024Apr}
\begin{equation}
 X(t) = \int_0^t dt' \sum_{\alpha,j} x_{\alpha j} \tr{\hat{L}_{\alpha, j}^\dag(t')\hat{L}_{\alpha, j}(t') \hat{\rho}_\MRS(t')}, 
\end{equation}
where $x_{\alpha j}$ are the weights with which each jump operator contributes to the variable $X$ of interest. The dynamical activity, $\AGKSL$, is defined as the average number of jumps observed within a given time interval of duration $t$~\cite{Carollo2019Apr,Hasegawa2020Jul,VanVu2022Apr}. Similarly to the classical counterpart, it is the counting observable where all of the weights are set to one, $x_{\alpha j} =1$~\cite{Landi2024Apr}. Expressed as the time-integral over the activity rate $\KGKSL$, it is hence given by
\begin{eqnarray}\label{eq: dynamical activity GKSL}
 \AGKSL(t) & = & \int_0^{t} dt' \KGKSL(t'),\\
\label{eq: activity rate GKSL}
 \KGKSL(t) & = & \sum_{\alpha,j}\tr{\hat{L}^\dag _{\alpha, j}(t)\hat{L}_{\alpha, j}(t) \hat{\rho}_\MRS(t)}.
\end{eqnarray}
We point out that this definition of activity is a quantity that is directly measurable, as one can, in principle, continuously monitor all jumps~\cite{Landi2024Apr}.

\subsubsection{Correlator-based dynamical activity in quantum transport}\label{sec:correlator_activity}
Defining a dynamical activity for quantum systems beyond weak coupling is an ongoing research topic. Recently, a definition for a quantum-transport setting has been introduced in Ref.~\cite{Blasi2026Jul}, based on the symmetrized correlator of the system-bath couplings. Concretely, for an open quantum system described by the total Hamiltonian \begin{equation}
 \hat{H} = \hat{H}_\MRS + \sum_{\alpha} \left(\hat{H}_\alpha + \hat{V}_\alpha\right) ,
\end{equation}
with $\hat{H}_\alpha$ and $\hat{V}_\alpha$ the bath and coupling Hamiltonians, respectively, 
 the quantum dynamical activity rate with respect to reservoir $\alpha$ was defined as the symmetrized correlator 
\begin{equation}\label{eq: corr based K}
 \mathcal{K}^{VV}_\alpha(t):=\frac{1}{2\hbar^2} \int_{-t}^td\tau \langle\langle \{\hat{V}^\mathrm{H}_\alpha(t) ,\hat{V}^\mathrm{H}_\alpha(t+\tau)\} \rangle\rangle.
\end{equation}
Here, $\hat{V}^\mathrm{H}_\alpha(t)$ is the coupling in the Heisenberg picture, $\langle\langle \hat{X} \hat{Y}\rangle\rangle := \langle\hat{X} \hat{Y}\rangle -\langle\hat{X}\rangle\langle \hat{Y}\rangle$ and $\langle \hat{X}\rangle :=\tr{ \hat{X}\hat{\rho}(t_0)}$ denote expectation values with respect to the initial density matrix. In Ref.~\cite{Blasi2026Jul}, Eq.~\eqref{eq: corr based K} was investigated for noninteracting electronic transport and shown to recover the GKSL activity rate~\eqref{eq: activity rate GKSL} in the weak-coupling regime for a general quadratic Hamiltonian. In Appendix~\ref{app: weak coupling correlator} we show that $\mathcal{K}^{VV}_\alpha(t)$ recovers the weak-coupling activity more generally. Moreover, as shown in Ref.~\cite{Palmqvist2026,Palmqvist2026Jul}, it is possible to make an information geometric interpretation of Eq.~\eqref{eq: corr based K} and connect it to the earlier definitions of quantum dynamical activity based on quantum Fisher information used in Refs.~\cite{Hasegawa2023May,Nishiyama2024Apr}.

Recently, a susceptibility kinetic uncertainty relation (S-KUR) has been found as a generalization of the classical KUR for general quantum transport settings~\cite{Palmqvist2026Jul}. The S-KUR not only provides a generally valid bound for quantum systems, but it also allows for bounding the precision by the activity~\eqref{eq: corr based K} of a single reservoir, in contrast to the standard classical KUR, which instead uses the total activity.
The S-KUR for steady-state transport of an observable $\hat{X}$ takes the form~\cite{Palmqvist2026Jul}
\begin{equation}\label{eq:SKUR}
\begin{aligned}
 \mathscr{P}^{(X)}_\alpha &:= \frac{\left(\mathcal{J}^{(X)}_\alpha\right)^2}{S^{(X)}_\alpha} \leq \mathcal{K}^\mathrm{lim}_\alpha =\mathcal{K}^{VV}_\alpha,
 \\
 \mathcal{K}^\mathrm{lim}_\alpha (t)&:= \frac{1}{\hbar^2} \int_{0}^td\tau \langle\langle \{\hat{V}^\mathrm{H}_\alpha(t) ,\hat{V}^\mathrm{H}_\alpha(\tau)\} \rangle\rangle
\end{aligned}
\end{equation}
where 
\begin{equation}\label{eq:response_current}
 \begin{aligned}
 \mathcal{J}^{(X)}_\alpha &= \frac{1}{2} (I^{(X)}_\alpha +\dot M^{(X)}_\alpha ),\\
 \dot M_\alpha^{(X)} &= \partial_\theta \left( \frac{I^{(X)}_\alpha(\theta)}{\theta}\right)\Bigg|_{\theta=1}.
 \end{aligned}
\end{equation}
Here, $\dot M^{(X)}_\alpha$ is a susceptibility of the current to a change in the coupling strength to reservoir $\alpha$ parametrized by $\theta$. Crucially, the susceptibility $\dot M^{(X)}$ is measurable whenever the coupling strength is tunable in experiment~\cite{Palmqvist2026Jul}. Notice that $\mathcal{K}^\mathrm{lim}_\alpha (t) =\mathcal{K}^{VV}_\alpha (t)$ under the assumption of time-translation invariance~\cite{Palmqvist2026Jul}.

However, the correlations of Eq.~\eqref{eq: corr based K} might be difficult to measure beyond the weak-coupling regime, and bounds in terms of more easily accessible observables are therefore also of interest.

\subsubsection{Dynamical-activity substitutes motivated by quantum-transport observables}\label{sec:phenomenological}
A complementary approach to establishing precision bounds is by identifying how measurable (transport) quantities constrain the precision of arbitrary currents. In Refs.~\cite{Palmqvist2025Oct,Palmqvist2025Jul}, it was shown that the particle-current noise, or more specifically the particle-current auto-correlations,
\begin{equation}
 S_{\alpha}^{(N)}=\int_{-\infty}^{\infty} dt \langle\langle \hat{I}_\alpha^{(N)}(t)\hat{I}_\alpha^{(N)}(0)\rangle\rangle,\label{eq:SN}
\end{equation}
can take the role of a local dynamical activity, replacing $\mathcal{K}$ in Eq.~\eqref{eq:KUR_intro}, as long as particle-particle interactions are weak or, in other words, the system Hamiltonian is bilinear in the creation/annihilation operators. Here, $\hat{I}_\alpha^{(N)}(t)$ is the particle current operator. Indeed, in the close-to-equilibrium or low-transparency regimes, the particle-current
noise is given by a sum of single-particle transfer rates
\begin{equation}
 S_{\alpha}^{(N)}=\Gamma_{\alpha,\rightarrow}+\Gamma_{\alpha,\leftarrow}\ .
\end{equation}
It thereby counts---in the spirit of a standard, classical activity---all processes of excitation transfers into or out of reservoir $\alpha$. In contrast to the above-introduced dynamical activities, it does, however, not account for back-reflection processes. 

The approach of replacing the dynamical activity by the particle-current noise is particularly appealing when considering thermodynamic inference, based on a TKUR, where a tighter bound on entropy production compared to the standard TUR can be formulated without needing to introduce any additional quantities for bosonic transport in linear systems~\cite{Palmqvist2025Jul}. 

Indeed, for linear systems, the particle-current noise provides a valid bound on precision; it is, however, only a tight bound as long as the system is close to equilibrium or has low transparency. To maintain a tight bound, a function of the particle-current noise involving even the average current and the transmission bandwidth has been found~\cite{Palmqvist2025Oct,Palmqvist2025Jul}.

\section{Bosonic transport in harmonic systems}\label{sec:HO}

\subsection{Energy transport in harmonic-oscillator network}
To model steady-state bosonic transport, we consider a harmonic network of arbitrary dimension, described by the Hamiltonian~\cite{Wang2014Dec}
\begin{equation}\label{eq: hamiltonian HOSC}
 \hat{H} = \frac{1}{2} \hat{p}^T \hat{p} + \frac{1}{2} \hat{u}^T \mathbf{K} \hat{u}\ .
\end{equation}
Here we denote the rescaled positions of the oscillators by $\hat{u}_i=\sqrt{m}_i\hat{x}_i$, where $\hat{u}$ and $\hat{p}$ are column vectors containing all the positions and momenta of the network. The spring-constant matrix $\mathbf{K}$ is symmetric and positive definite. We partition the network into a central region S and baths, which we label by Greek letters $\alpha,\beta$. Assuming that there is no reservoir-reservoir interaction, the Hamiltonian can be partitioned as~\cite{Wang2014Dec} 
\begin{subequations}
 \begin{eqnarray}
 \hat{H} &=& \frac{1}{2} \hat{p}_\MRS^T \hat{p}_\MRS + \frac{1}{2} \hat{u}_\MRS^T \mathbf{K}_\MRS \hat{u}_\MRS \nonumber\\
 &&+\sum_\alpha (\hat{H}_\alpha + \hat{u}^T_\alpha \MBV_{\alpha \MRS} \hat{u}_\MRS), \\
 \hat{H}_\alpha &= &\frac{1}{2}\hat{p}_\alpha^T \hat{p}_\alpha + \frac{1}{2} \hat{u}_\alpha^T \mathbf{K}_\alpha \hat{u}_\alpha ,
\end{eqnarray}
\end{subequations}
where $\MBK_\alpha$ is the spring-constant matrix of bath $\alpha$, uncoupled from the rest of the network, and $\MBV_{\alpha \MRS} = (\MBV_{\MRS \alpha })^T$ is the part of the spring constant matrix which describes the coupling between the positions of bath $\alpha$ and the central region S.

The transmission between baths via the central region, important for the transport analysis, is characterized by a transmission matrix and transmission function~\cite{Caroli1971Jun},
\begin{subequations}
 \begin{eqnarray}\label{eq: def Tab matrix}
 \mathbf{T}_{\alpha \beta}(\omega) & = & \boldsymbol{\Gamma}_\alpha (\omega)\boldsymbol{G}_{\MRS\MRS}^r (\omega)\boldsymbol{\Gamma}_\beta(\omega) \boldsymbol{G}_{\MRS\MRS}^a(\omega),\\
 \label{eq: def Tab}
 \mathcal{T}_{\alpha \beta}(\omega) & = & \tr{\mathbf{T}_{\alpha \beta}(\omega)},
\end{eqnarray}
\end{subequations} 
where $\boldsymbol{G}_{\MRS\MRS}^{r,a}(\omega)$ are matrices containing all the retarded and advanced Green's functions of the central region, and $\mathbf{\Gamma}_\alpha(\omega) $ is the spectral function of bath $\alpha$, details are given in Appendix~\ref{app: deriv gen K}. The occupation of bath $\alpha$ is given by the Bose-Einstein distribution $n_\alpha(\omega)= 1/(e^{\hbar\omega/\kB T_\alpha}-1)$.

We are interested in the energy current out of one of the baths and in the precision of this energy current. Since bounds on the energy-current precision can be formulated in terms of the particle-current noise, see Sec.~\ref{sec:phenomenological}, we here also calculate particle transport. The particle current should in this type of system rather be understood as ``excitation current'' and it is typically not conserved; however in particular its fluctuations are of interest here to characterise how ``active'' the system is and to establish precision bounds. We evaluate the excitation current in the eigenbasis of the uncoupled bath $\alpha$, where currents are chosen to be detected.

The energy and excitations transferred out of reservoir $\alpha$ at time $t$ are given by 
\begin{subequations}
 \begin{eqnarray}
 \Delta \hat{E}_\alpha(t) & = & \int^t_0 dt' \hat{I}^{(E)}_\alpha(t') = \hat{H}_\alpha(0)- \hat{H}^\mathrm{H}_\alpha(t),\\
 \Delta \hat{N}_\alpha(t) & = & \int^t_0 dt' \hat{I}^{(N)}_\alpha(t') = \hat{N}_\alpha(0)- \hat{N}^\mathrm{H}_\alpha(t),
\end{eqnarray}
\end{subequations}
where $\hat{H}^\mathrm{H}_\alpha(t)$ is the bath Hamiltonian and $\hat{N}^\mathrm{H}_\alpha(t)$ is the number operator, counting the excitations in each eigenstate of the bath. The superscript H denotes that the operators are written in the Heisenberg picture. The current operators are given by
\begin{subequations}
 \begin{eqnarray}
 \hat{I}^{(E)}_\alpha(t)& := & - \frac{d}{dt} \hat{H}^\mathrm{H}_\alpha(t),\\
 \hat{I}^{(N)}_\alpha(t) & := & - \frac{d }{dt} \hat{N}^\mathrm{H}_\alpha(t).
\end{eqnarray}
\end{subequations}
 A convenient way to calculate steady-state excitation and energy currents and zero-frequency fluctuations in this type of system is by using the full counting statistics (FCS) combined with the nonequilibrium Green's functions approach~\cite{PismaZhETF.58.225,levitov_electron_1996,Nazarov2003Oct,Esposito2009Dec,Lesovik2011Oct,Agarwalla2012May,Tang2014Nov}.

Since $\hat{H}_\alpha$, $\hat{N}_\alpha$ admit a joint eigenbasis $\ket{\epsilon,n}$ (excitation energies and numbers in the eigenmodes of reservoir $\alpha$), we can treat the FCS of energy and excitations with a single generating function using the two-point measurement scheme~\cite{Esposito2009Dec}. We define the probability to detect the energy and excitation number $\epsilon_0,n_0$ in reservoir $\alpha$ at time $t'=0$ and $\epsilon_t,n_t$ in reservoir $\alpha$ at time $t'=t>0$,
\begin{equation}
\begin{aligned}
 \label{eq:def_TPM}P(\Delta\epsilon , \Delta n) = \sum_{n_t,n_0,\epsilon_t,\epsilon_0}& \delta(\Delta \epsilon - (\epsilon_t -\epsilon_0) ) \delta(\Delta n - (n_t -n_0) ) \\
 &\times P(\epsilon_t,n_t,\epsilon_0,n_0).
\end{aligned}
\end{equation}
with the two-point-measurement probability
\begin{equation}
 P(\epsilon_t,n_t,\epsilon_0,n_0) =\tr{\hat{\Pi}_{t} \hat{U}(t,0) \hat{\Pi}_{0} \hat{\rho}(0) \hat{\Pi}_{0} \hat{U}^\dag(t,0) \hat{\Pi}_{t}}.
\end{equation}
 Here, $\hat{U}(t,0) = T_+\exp{\{-i \int _0^tdt' \hat{H}(t')/\hbar \}}$ is the time-evolution operator, $\hat{\Pi}_{t} = \ket{\epsilon_t,n_t} \bra{\epsilon_t, n_t}$ the projector on states in reservoir $\alpha$, and $\hat{\rho}(0)$ is the initial density operator of the total system.
Based on this, the moment-generating function is given by 
\begin{equation}
\begin{aligned}
 \mathcal{Z}_\alpha(\xi_E, \xi_N) &:= \iint d\Delta\epsilon\; d\Delta n \;e^{i \xi_E \Delta \epsilon} e^{i \xi_N \Delta n} P(\Delta\epsilon , \Delta n) \\
 &=\left\langle e^{i( \xi_E \hat{H}_\alpha+ \xi_N \hat{N}_\alpha )} e^{-i( \xi_E \hat{H}^\mathrm{H}_\alpha(t)+ \xi_N \hat{N}^\mathrm{H}_\alpha(t) )} \right\rangle.
\end{aligned}
\end{equation}
Here, we have used the definition given in Eq.~\eqref{eq:def_TPM} and we have assumed that the initial density matrix is diagonal in the number basis, such that~\cite{Esposito2009Dec}
\begin{equation}
 \begin{aligned}
 &\sum_{n_0} e^{-i(\xi_E \epsilon_0 +\xi_N n_0)} \hat{\Pi}_{n_0} \hat{\rho}(0)\hat{\Pi}_{n_0}= \\ &e^{ -i(\xi_E \hat{H}_\alpha +\xi_N \hat{N}_\alpha)/2} \hat{\rho}(0) e^{ -i(\xi_E \hat{H}_\alpha +\xi_N \hat{N}_\alpha)/2}\ .
 \end{aligned}
\end{equation}
In the steady-state limit, under the assumptions of the bath being large and thermal, one can exploit the fluctuation-dissipation relation to write the corresponding steady-state cumulant generating function (CGF) $\chi_\alpha(\xi_E,\xi_N):= \lim_{t\rightarrow\infty} \frac{1}{t}\ln \mathcal{Z}_\alpha(\xi_E,\xi_N)$ as~\cite{Agarwalla2012May}
\begin{equation}\label{eq: CGF HOSC}
 \begin{aligned}
 &\chi_\alpha(\xi_E,\xi_N)=
 - \int^\infty_{-\infty} \frac{d\omega}{4 \pi} \ln\det\Big\{\mathbb{I} - \sum_{\beta\neq\alpha} \mathbf{T}_{\alpha \beta}\times\\ &\times \left[(e^{i (\xi_E \hbar \omega + \xi_N)} -1) F_{\alpha \beta} +(e^{-i( \xi_E \hbar \omega+\xi_N)} -1) F_{ \beta\alpha} \right] \Big\},
 \end{aligned}
\end{equation}
with the transmission matrix given in Eq.~\eqref{eq: def Tab matrix}. Here, we defined $F_{\alpha \beta}(\omega)=n_\alpha(\omega)(1+n_\beta(\omega))$ and suppressed the $\omega$ arguments for brevity.

For the average currents, one finds 
\begin{subequations}
\begin{eqnarray}
 I^{(N)}_\alpha &=&\frac{\partial\chi_\alpha(\xi_E,\xi_N)}{\partial(i\xi_N)}|_{\xi_E,\xi_N=0} \nonumber\\ 
 &=& \int^\infty_0\frac{d\omega}{2\pi} \sum_\beta \mathcal{T}_{\alpha \beta}(n_\alpha-n_\beta),\label{eq:part_curr}\\
 I^{(E)}_\alpha &=&\frac{\partial\chi_\alpha(\xi_E,\xi_N)}{\partial(i\xi_E)}|_{\xi_E,\xi_N=0}\nonumber \\
 &=& \int^\infty_0\frac{d\omega}{2\pi} \sum_\beta \mathcal{T}_{\alpha \beta}\hbar \omega (n_\alpha-n_\beta) .\label{eq:en_curr}
\end{eqnarray}
\end{subequations}
Furthermore, the zero-frequency noise of excitation and energy currents, $S^{(N)}_\alpha = \partial^2 \chi_\alpha(\xi_E,\xi_N)/\partial^2(i\xi_N)|_{\xi_E,\xi_N=0}$ and $S^{(E)}_\alpha = \partial^2 \chi_\alpha(\xi_E,\xi_N)/\partial^2(i\xi_E)|_{\xi_E,\xi_N=0}$ are found as\footnote{ Note that Equation~(79) of Ref.~\cite{Agarwalla2012May} is equivalent to these equations only due to the fact that $\mathbf{K}$ is symmetric and hence also the transmission functions \eqref{eq: def Tab} are. }
\begin{subequations}
\begin{eqnarray}
 S^{(N)}_\alpha 
 &=& \int^\infty_0\frac{d\omega}{2\pi} \left(\sum_{\beta\neq \alpha} \mathcal{T}_{\alpha \beta} \left[F_{\alpha \beta}+ F_{ \beta\alpha}\right]\right.\nonumber\\
 &&+ \left.\mathrm{tr}\Big\{\Big[\sum_\beta \mathbf{T}_{\alpha \beta} (n_\alpha -n_\beta) \Big]^2\Big\}\label{eq:part_fluct}\right) ,\\
 S^{(E)}_\alpha 
 &=& \int^\infty_0\frac{d\omega}{2\pi} \left(\sum_{\beta\neq \alpha}(\hbar \omega)^2 \mathcal{T}_{\alpha \beta}\left[F_{\alpha \beta}+ F_{ \beta\alpha}\right]\right.\nonumber\\
 &&+\left.\mathrm{tr}\Big\{\Big[\hbar \omega\sum_\beta \mathbf{T}_{\alpha \beta} (n_\alpha -n_\beta) \Big]^2\Big\}\right) .\label{eq:en_fluct}
\end{eqnarray} 
\end{subequations}
The expression for the noise can be split into two parts where the first lines of these two expressions are often referred to as classical noise, $S^{(N),\mathrm{cl}}_\alpha$ and $S^{(E),\mathrm{cl}}_\alpha$, and the second lines as quantum parts of the noise, $S^{(N),\mathrm{qu}}_\alpha$ and $S^{(E),\mathrm{qu}}_\alpha$. In the following, we use the excitation-current fluctuations~\eqref{eq:part_fluct} to establish activity-measures, and the energy current~\eqref{eq:en_curr} to define entropy production in order to bound the energy-current fluctuations.

\subsection{Entropy production and activities}

In addition to the transport currents and their fluctuations, we now evaluate the entropy and different notions of activity for the harmonic-oscillator network, which are required to investigate the thermodynamic and kinetic uncertainty relations introduced in Sec.~\ref{sec:intro_tradeoffs}. 

To evaluate the entropy production, we use Eq.~\eqref{eq:entropy_clausius} together with~\eqref{eq:en_curr}, and find~\cite{Brandner2025Feb,Palmqvist2025Jul}
\begin{equation}
 \sigma = \kB \sum_{\alpha\beta} \int^\infty_0 \frac{d\omega}{2\pi} \mathcal{T}_{\alpha \beta}\ln\left[ \frac{F_{\alpha \beta}}{F_{\beta\alpha }}\right]\frac{(F_{\alpha \beta}-F_{\beta\alpha })}{2},
\end{equation}
if the transmission function is symmetric $\T_{\alpha \beta}(\omega)=\T_{\beta\alpha }(\omega)$ (which is always the case here).

In addition, we need a notion of activity. We here focus on the regime of \textit{strong coupling}, where the Lindblad equation does not apply, and therefore follow the strategies of Secs.~\ref{sec:correlator_activity} and \ref{sec:phenomenological}. Concretely, the correlator-based activity rate [Eq.~\eqref{eq: corr based K}] in the steady state is found to take the form
\begin{subequations}\label{eq: gen K HOSC}
\begin{equation}
 \begin{aligned}
 \mathcal{K}^{VV}_\alpha&= \lim_{t\rightarrow\infty} \mathcal{K}^{VV}_\alpha(t) =\mathcal{K}^\text{cross}_\alpha +\mathcal{K}^\text{auto}_\alpha ,
 \end{aligned}
\end{equation}
where
\begin{align}
 \mathcal{K}^\text{cross}_\alpha := \int^\infty_0\frac{d\omega}{2\pi} \sum_{\beta\neq \alpha} \mathcal{T}_{\alpha \beta} (F_{\alpha \beta}+ F_{ \beta\alpha})=S^{(N),\mathrm{cl}}_\alpha 
\label{eq:K_cross_HO}\\
 \mathcal{K}^\text{auto}_\alpha := \int^\infty_{0} \frac{d\omega}{2\pi} \mathrm{tr}\Big\{4\mathbf{T}_{\alpha \alpha}-\Big[\sum_\beta \mathbf{T}_{\alpha\beta}\Big]^2\Big\} F_{\alpha \alpha}.\label{eq:K_auto_HO}
\end{align}
\end{subequations}
The expression for Eq.~\eqref{eq: gen K HOSC} is analogous to the fermionic result of Ref.~\cite{Blasi2026Jul} which is recovered by the replacement $F_{\alpha\beta}=n_\alpha (1+n_\beta) \mapsto f_\alpha (1-f_\beta)$, where $f_\alpha(E)$ is the Fermi-Dirac distribution. For a full derivation of Eq.~\eqref{eq: gen K HOSC} see Appendix~\ref{app: deriv gen K}. 

Following the derivation of Refs.~\cite{Palmqvist2025Oct,Palmqvist2025Jul}, one finds
\begin{equation}
 \mathcal{P}^{(E)}_\alpha \leq \mathcal{K}^\mathrm{cross}_\alpha \leq S^{(N)}_{\alpha},
\end{equation}
and 
\begin{equation}
 \sum_\alpha C(S^{(N)}_{\alpha} ,\mathcal{P}^{(E)}_\alpha )\leq\sum_\alpha C(\mathcal{K}_\alpha^\mathrm{cross} ,\mathcal{P}^{(E)}_\alpha ) \leq \frac{\sigma}{\kB}
\end{equation}
or
\begin{equation}
 2C(S^{(N)}_{\alpha} ,\mathcal{P}^{(E)}_\alpha ) \leq 2C(\mathcal{K}_\alpha^\mathrm{cross} ,\mathcal{P}^{(E)}_\alpha ) \leq \frac{\sigma}{\kB}.
\end{equation}

\subsection{Trade-off relations for precision}
To explore the precision limits based on the activities and the entropy production introduced above in a concrete example, we consider the case when the central region consists of a single harmonic oscillator with frequency $\omega_0$ connected to left and right reservoirs, L and R. The two reservoirs are characterized by different temperatures $T_\mathrm{L} =\bar{T} +\Delta T/2$ and $T_\mathrm{R} =\bar{T} -\Delta T/2$. The transmission function is given by~\cite{Saito2007Oct,Ojanen2008,Agarwalla2012May,Segal2016May,Agarwalla2017Apr}
\begin{equation}
 \mathcal{T}_{\alpha \beta}(\omega)=\frac{ \Gamma_\alpha (\omega)\Gamma_\beta (\omega) }{(\omega^2-\omega_0^2)^2 + \frac{1}{4} [\Gamma_\mathrm{L}(\omega) +\Gamma_\mathrm{R}(\omega)]^2}.\label{eq:transmission_HO}
\end{equation}
Furthermore, we assume ohmic spectral functions with an exponential cutoff~\cite{Weiss2012}, 
\begin{equation}
 \Gamma_\alpha(\omega) = 2 g_\alpha \omega e^{-\omega /\omega_c}.
\end{equation}
For this limit, we analyze bounds on the precision of the energy current $\mathcal{P}^{(E)}_\mathrm{L} = (I^{(E)}_\mathrm{L})^2 /S^{(E)}_\mathrm{L}$. Note that, while we focus on energy-current precision here, in principle, even the precision of the excitation current, $\mathcal{P}^{(N)}_\mathrm{L} = (I^{(N)}_\mathrm{L})^2 /S^{(N)}_\mathrm{L}$ could be considered. 

\subsubsection{TUR and local KURs}
\begin{figure}[b!]
 \centering
 \includegraphics[width=3.3in]{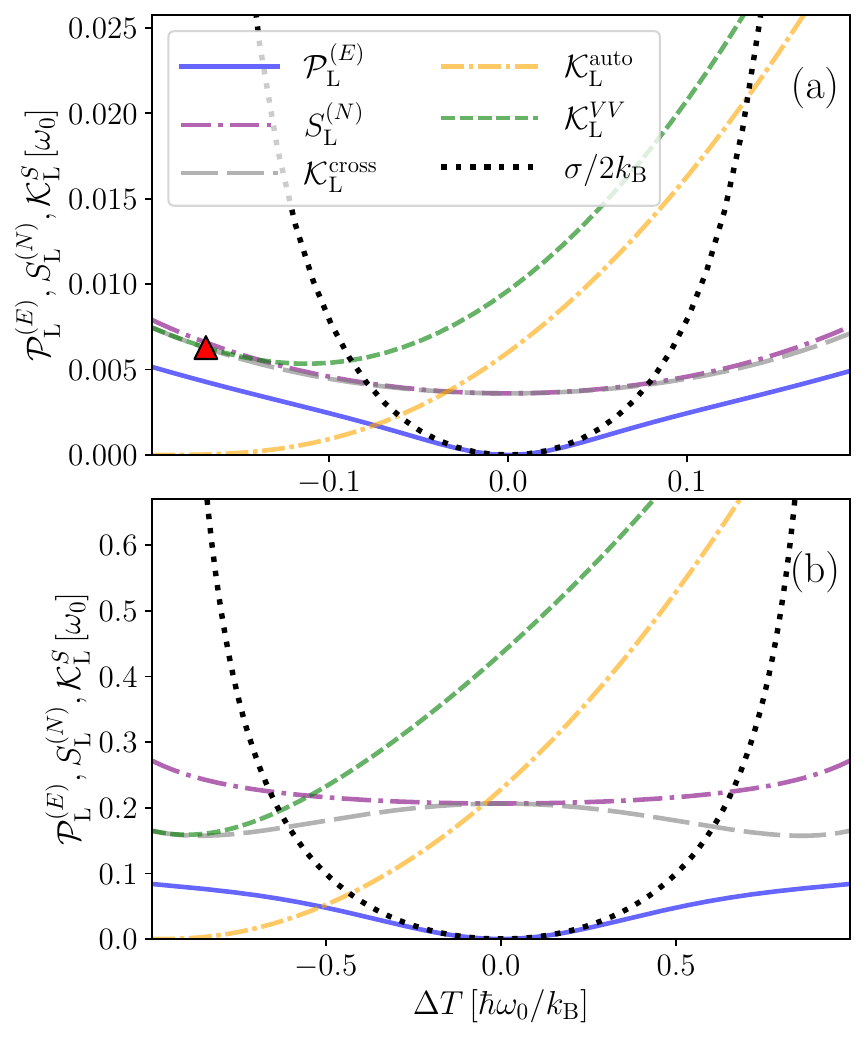}
\caption{Harmonic-oscillator system: Precision in energy current $\mathcal{P}^{(E)}_\mathrm{L}$, the current noise $S^{(N)}_\mathrm{L}$, contributions to the correlator-based activity $\mathcal{K}^{VV}_\mathrm{L}= \mathcal{K}^\mathrm{cross}_\mathrm{L} +\mathcal{K}^\mathrm{auto}_\mathrm{L}$, and entropy production $\sigma$ as functions of temperature bias $\Delta T $. In~(a) the average temperature is $\bar{T} = 0.1 \hbar\omega_0 /\kB$ and in~(b) $\bar{T} = \hbar\omega_0/\kB$. In both panels, the coupling strengths are set to $g_\alpha =\omega_0$ and the cutoff frequency $\omega_c =10 \omega_0$. The red triangle indicates one of the points appearing in Fig.~\ref{fig: Hosc comp sampling}. }
 \label{fig: Hosc comp KUR}
\end{figure}
In Fig.~\ref{fig: Hosc comp KUR}, we show the results for energy-current precision together with the \textit{full} entropy production and with different definitions of the \textit{local} activity and their constituents as functions of temperature bias $\Delta T$.
Indeed, for noninteracting systems, as the local network considered here, local versions of the bounds introduced in Sec.~\ref{sec:intro_tradeoffs} hold and---since obviously $\mathcal{K}_\mathrm{L} \leq \mathcal{K}_\mathrm{L} +\mathcal{K}_\mathrm{R}$---turn out to be tighter than the global counterparts~\cite{Palmqvist2025Oct}.

At $\Delta T =0$, the precision is always zero since there is no average current. Close to equilibrium, the entropy production $\sigma$ gives a tight constraint on precision through the TUR. By contrast, the KUR is in general less predictive close to equilibrium, but tighter than the constraint given by the TUR far away from equilibrium, as expected. 

We now compare the KUR constraints based on the different activity concepts introduced in Sec.~\ref{sec:dynamical_activity}. We first notice that the tightest bound is provided by the activity contribution $\mathcal{K}^\mathrm{cross}_\mathrm{L}=S^{(N),\mathrm{cl}}_\mathrm{L}$ and it is closest to the actual energy-current precision in the low-temperature regime (panel (a)), but never actually reaches it. This is explained by the following effects: The bound given by $\mathcal{K}^\mathrm{cross}_\mathrm{L}=S^{(N),\mathrm{cl}}_\mathrm{L}$ is expected to be tight when the difference between left and right bath occupation is maximal~\cite{Palmqvist2025Oct}. Therefore, it is most predictive when one of the contacts is ``empty'' (which is best approached close to zero-temperature) within the transport window. Here, however, the transmission of the harmonic oscillator network---and hence the transport window---is broadened, see Eq.~\eqref{eq:transmission_HO}, such that there are always transport contributions in regions where the occupations of the two contacts are similar to each other, even when the temperature of one of the contacts tends towards zero. The contribution $\mathcal{K}^\mathrm{cross}_\mathrm{L}$ in turn is limited both by the correlator-based activity and by the excitation-current noise 
$\mathcal{P}^{(x)}_\mathrm{L} \leq \mathcal{K}^\mathrm{cross}_\mathrm{L}\leq\mathcal{K}^{VV}_\mathrm{L}, S^{(N)}_\mathrm{L}$, as expected~\cite{Palmqvist2025Oct,Blasi2026Jul}. Close to equilibrium and in the low-temperature limit, the bound provided by the full excitation-current noise basically coincides with $\mathcal{K}^\mathrm{cross}_\mathrm{L}$; instead, away from equilibrium, we have $S^{(N)}_\mathrm{L} > S^{(N),\mathrm{cl}}_\mathrm{L}$ due to $ S^{(N),\mathrm{qu}}_\mathrm{L}>0$. This is because, the larger the temperature and the temperature bias, the more the nonequilibrium contribution, $n_\alpha-n_\beta$ underlying the quantum contribution to the noise becomes significant. Also, the excitation-current noise $S^{(N)}_\mathrm{L}$ is symmetric with respect to changing the sign $\Delta T$, just as $\mathcal{K}^\mathrm{cross}_\mathrm{L}$. 
 By contrast, $\mathcal{K}_\mathrm{L}^{VV}$ is asymmetric with respect to the temperature bias. This is due to the integrand in $\mathcal{K}^\mathrm{auto}_\mathrm{L}$ containing $F_{\mathrm{L} \mathrm{L}}(\omega)$ which causes $\mathcal{K}^\mathrm{auto}_\mathrm{L}$ to become small when $T_\mathrm{L}$ is small, and large when $T_\mathrm{L}$ is large. In particular, when $T_\mathrm{L}\rightarrow0$, one finds $\mathcal{K}^\mathrm{auto}_\mathrm{L}\rightarrow 0$. Note that this asymmetry is far less pronounced in fermionic systems~\cite{Blasi2026Jul}, where the factor $F_{\mathrm{L}\mathrm{L}}$ is bounded by 1. 
Thus, depending on the temperature of contact $\mathrm{L}$ and on the temperature bias, either the particle current noise $S^{(N)}_\mathrm{L}$ or the correlator-based definition of dynamical activity will provide a tighter constraint on precision.

\subsubsection{TKUR and entropy-inference bound}
\begin{figure}[t!]
 \centering
 \includegraphics[width=3.3in]{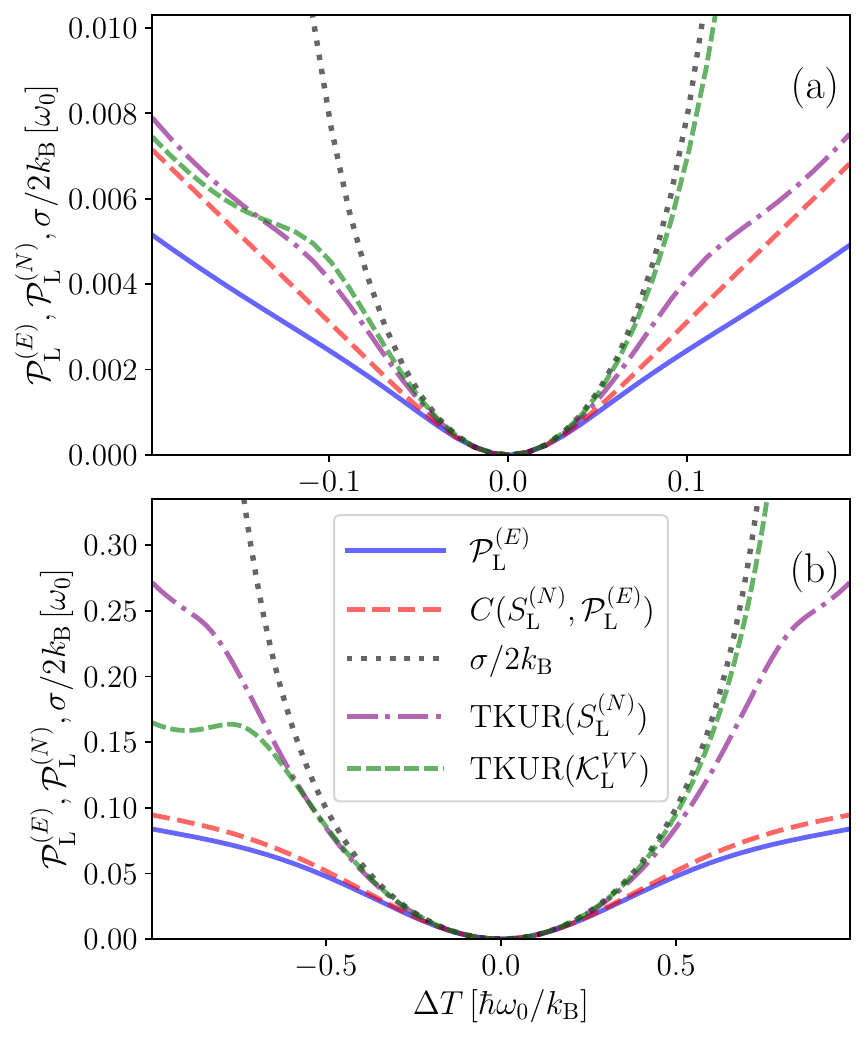}
\caption{Harmonic-oscillator system: Precision in energy current $\mathcal{P}^{(E)}_\mathrm{L}$, TKURs using particle current noise and the correlator-based definition of activity to estimate single-particle transfer rates and entropy production $\sigma$ as functions of temperature bias $\Delta T =T_\mathrm{L}-T_\mathrm{R}$. In addition, the improved inference-bound on entropy production of~\eqref{eq:bosonic linear bound} is shown. In panel~(a) the average temperature $\bar{T}= (T_\mathrm{L}+T_\mathrm{R})/2 = 0.1 \hbar\omega_0 /\kB$ and in panel~(b) $\bar{T} = \hbar\omega_0/\kB$. In both panels, the coupling strengths are set to $g_\alpha =\omega_0$ and the cutoff frequency $\omega_c =10 \omega_0$. }
 \label{fig: Hosc comp TKUR}
\end{figure}

In Fig.~\ref{fig: Hosc comp TKUR}, we analyze bounds on the energy-current precision given by the thermokinetic uncertainty relation (TKUR), see Eq.~\eqref{eq:TKUR_intro}, again using the different activities introduced above. In general, we see that the TKUR bounds are close to the TUR for small biases and provide better bounds than the TUR far away from equilibrium. However, also here, the bounds are not tight. Due to the previously described asymmetry in the correlator-based activity, the TKUR using $\mathcal{K}_\mathrm{L}^{VV}$ remains close to the TUR when the left lead is hot and it provides a tighter bound than the TKUR using the excitation-current noise as activity substitute when the left contact is cold. This is particularly evident when the average temperature is large (of the order of the level spacing, see panel (b)). 

Importantly, Fig.~\ref{fig: Hosc comp TKUR} also shows that the log-bound of Eq.~\eqref{eq:bosonic linear bound} constitutes a lower bound to the produced entropy. However, in this case, it is only a moderate improvement for inferring the entropy-production rate compared to the TUR itself, which is here due to the fact that the energy current displays significant noise. If the transmission function has sharp features, improvements can be made by including a bandwidth over which transport occurs in the function $C$~\cite{Palmqvist2025Jul}.

\subsubsection{Susceptibility-KUR}
\begin{figure}[t!]
 \centering
 \includegraphics[width=3.3in]{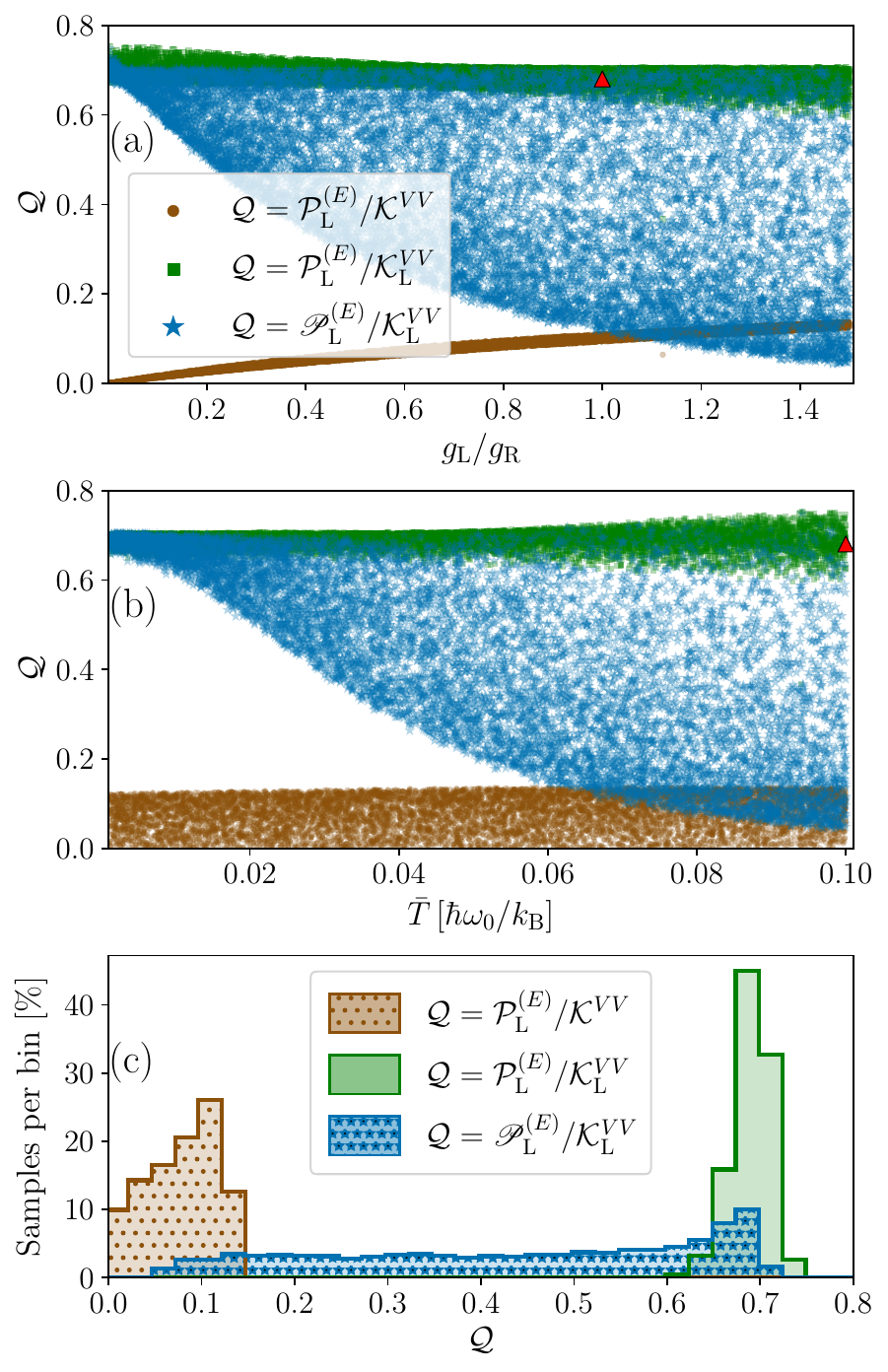}
\caption{Harmonic-oscillator system: Saturation of local KUR (Eq.~\eqref{eq:local KUR} green squares and histogram), S-KUR (\eqref{eq:SKUR} blue stars and histogram) and classical KUR (Eq.~\eqref{eq:global KUR} brown dots and histogram). We show the ratios between precision and bound as function of coupling asymmetry in panel~(a) and as function of the average temperature in~(b) where the remaining parameters are uniformly sampled (10000 points) with $\kB\bar{T}/(\hbar \omega_0) \in [10^{-3},10^{-1} ]$, $\Delta T/\bar{T}\in [ -0.95,-0.7]$ and $g_\MRL/g_\MRR \in [10^{-3},1.5]$ and $\omega_c = 10\omega_0$. Panel~(c) shows the statistics of values obtained for the ratio between precision and bound. The red triangle indicates one of the points appearing in Fig.~\ref{fig: Hosc comp KUR}. }
 \label{fig: Hosc comp sampling}
\end{figure}

To consider the S-KUR for the harmonic oscillator, we rescale the coupling strength to the left reservoir by $\theta$, which is inherited by the energy current as a parametrization of the transmission function
\begin{equation}
 I^{(E)}_\MRL(\theta) =\int^\infty_0\frac{d\omega}{2\pi} \mathcal{T}_{\MRL\MRR}(\omega;\theta)\hbar \omega (n_\MRL(\omega)-n_\MRR(\omega)) 
\end{equation}
\begin{equation}
 \mathcal{T}_{\MRL\MRR}(\omega;\theta)=\frac{ \theta^2 \Gamma_\MRL (\omega)\Gamma_\MRR (\omega) }{(\omega^2-\omega_0^2)^2 + \frac{1}{4} [\theta^2\Gamma_\mathrm{L}(\omega) +\Gamma_\mathrm{R}(\omega)]^2},\label{eq:theta transmission_HO}
\end{equation}
and the response current S-KUR~\eqref{eq:SKUR} according to Eq.~\eqref{eq:response_current}. The physical current is recovered at $\theta=1$. We compare the saturation of the S-KUR with the bosonic local KUR and the standard KUR using the activity of both reservoirs
\begin{subequations}
\begin{eqnarray}
 \mathcal{P}^{(E)}_\MRL &\leq& \mathcal{K}^{VV}_\MRL, \label{eq:local KUR}\\
 \mathcal{P}^{(E)}_\MRL &\leq& \mathcal{K}^{VV} := \mathcal{K}^{VV}_\MRL +\mathcal{K}^{VV}_\MRR, \label{eq:global KUR}\\
 \mathscr{P}^{(E)}_\MRL &:=& (\mathcal{J}^{(E)}_\MRL)^2/S^{(E)}_\MRL \leq \mathcal{K}^{VV}_\MRL.
\end{eqnarray}
\end{subequations}
We show the results in Fig.~\ref{fig: Hosc comp sampling} by sampling values of $\bar{T}$, $\Delta T$, and the coupling asymmetry $g_\MRL/g_\MRR$. For the considered ranges, the local KUR reaches a maximum saturation of $0.752$, comparable to the S-KUR that reaches $0.749$, while the saturation of the standard KUR never exceeds $0.138$. Interestingly, the saturation of the S-KUR is much more sensitive to the parameters, in particular to the coupling asymmetry and average temperature, as analyzed in panels~(a) and~(b). The strong impact of the coupling asymmetry originates from the fact that the S-KUR is tight when the current response to changes in the left reservoir-coupling strength is large. This happens, e.g., when the left reservoir coupling acts as a bottleneck for the current flow. Furthermore, one might indeed expect the S-KUR to be particularly tight at low temperatures, when the system approaches a pure state~\cite{Palmqvist2026Jul}.

\section{Nonequilibrium Spin-Boson model (NESB)}\label{sec:NESB}

We now investigate how local precision bounds are impacted or violated by strong anharmonicity. A paradigmatic example is the nonequilibrium spin-boson model (NESB)~\cite{Leggett1987Jan,Segal2005PRL,Segal2006,Weiss2012,Saito2013,Wang2017Feb}, which can be viewed as the hard-core limit of a bosonic mode: as an on-site interaction or Kerr nonlinearity becomes very large, multiple occupation is suppressed, and only the states with zero and one excitation remain relevant. The resulting effective system is a two-level system (TLS) coupled to bosonic reservoirs. This makes the NESB a suitable model to test to which extent local bounds, which are valid in transport through linear bosonic systems, survive in the presence of strong nonlinearities. In its most common form~\cite{Leggett1987Jan}, the spin-boson Hamiltonian describes the dynamics of a fictitious particle tunneling between two localized sites, and longitudinally [transverse coupling in Eq.~\eqref{eq: ham NESB}] coupled to bosonic reservoirs. Here, we limit ourselves to the case of the symmetric NESB, namely in the absence of a transverse field, where the bias-splitting vanishes, yielding a purely transverse coupling in the TLS energy eigenbasis\footnote{Note that a nonzero bias splitting would not result in additional violations of the classical TUR, KUR or TKUR on top of the ones we discuss in the following.}. The Hamiltonian for the TLS interacting with two bosonic baths, L and R, is then given by 
\begin{equation}\label{eq: ham NESB}
 \hat{H} =\frac{\hbar \Delta}{2} \hat{\sigma}_z +\sum_{\alpha j} \hbar \omega_{\alpha,j} \hat{b}^\dag_{\alpha,j} \hat{b}_{\alpha,j} + \sum_{\alpha} \hat{\sigma}_x \hat{B}_\alpha, 
\end{equation}
where the collective bath operator of bath $\alpha$ has the form 
\begin{equation}
 \hat{B}_\alpha=\sum_{j} \hbar \lambda_{\alpha,j}(\hat{b}_{\alpha,j} +\hat{b}_{\alpha,j}^\dag) \;.
\end{equation}

\subsection{Energy transport, entropy production and activities}
In order to isolate nonlinear effects, we here focus on the weak-coupling limit where we can employ the standard GKSL approach, Eq.~\eqref{eq:GKSL}, with the following jump operators
\begin{equation}
 \hat{L}^+_{\alpha} := \sqrt{\gamma_\alpha n_\alpha} \hat{\sigma}^+,\quad \hat{L}^-_{\alpha} := \sqrt{\gamma_\alpha(1+ n_\alpha)} \hat{\sigma}^-,
\end{equation}
where the Bose-Einstein distributions are evaluated at the level splitting $\Delta$, $n_\alpha:=n_\alpha( \Delta)$ and $\gamma_\alpha=\sum_j |\lambda_{\alpha,j}|^2 \delta(\Delta - \omega_{\alpha,j})$ is the coupling strength between bath $\alpha$ and the TLS, and $\hat{\sigma}^+ := \ket{e}\bra{g}$, $\hat{\sigma}^- := \ket{g}\bra{e}$. The steady-state density matrix follows from solving the GKSL equation $\mathcal{L} \hat{\rho}_\MRS = 0$, see Eq.~\eqref{eq:GKSL}, with $\hat{\rho}_\MRS = \lim_{t\rightarrow\infty} \hat{\rho}_\MRS(t)$, finding~\cite{Segal2006}
\begin{equation}
 \begin{aligned}
 \hat{\rho}_\MRS = \left(
\begin{array}{cc}
 \frac{\gamma_\mathrm{L}n_\mathrm{L}+ \gamma_\mathrm{R}n_\mathrm{R}}{\Lambda} & 0 \\
 0 & \frac{\gamma_\mathrm{L}(1+n_\mathrm{L})+ \gamma_\mathrm{R}(1+n_\mathrm{R})}{\Lambda} \\
\end{array}
\right).
 \end{aligned}
\end{equation}
Here, we have introduced the decay rate $\Lambda = \sum_{\alpha = \mathrm{L},\mathrm{R}} \gamma_\alpha(1+2n_\alpha)$, which is the negative of the relevant non-zero eigenvalue of the Liouvillian $\mathcal{L}$. The steady-state energy and particle currents out of the left bath differ from each other only by an energy factor $\hbar\Delta$ in this simple model system. They are computed as
\begin{equation}
 \begin{aligned}
 I^{(N)}_\mathrm{L}=\frac{I^{(E)}_\mathrm{L}}{\hbar \Delta} &= \tr{\hat{\rho}_\MRS({\hat{L}^+_\mathrm{L}})^\dag \hat{L}^+_\mathrm{L}}- \tr{\hat{\rho}_\MRS({\hat{L}^-_\mathrm{L}})^\dag \hat{L}^-_\mathrm{L} } \\
 &=\frac{ \gamma_\mathrm{L}\gamma_\mathrm{R}(n_\mathrm{L}-n_\mathrm{R}) }{\Lambda}.
 \end{aligned}
\end{equation}
We now introduce the activity and entropy production rates, which are required for evaluating the precision bounds. Both the limiting activity~\cite{Palmqvist2026Jul} and the correlator-based activity~\cite{Blasi2026Jul} reduce to the standard notion of activity for GKSL dynamics, see Appendix~\ref{app: weak coupling correlator},
\begin{equation}
 \begin{aligned}
 \mathcal{K}^{VV}_\alpha &= \frac{1}{2\hbar^2} \int^\infty_{-\infty} dt' \langle \langle \{\hat{V}_{\alpha}(t),\hat{V}_{\alpha}(t+t') \}\rangle \rangle\\
 &= \tr{\hat{\rho}_\MRS (\hat{L}^+_\alpha)^\dag \hat{L}^+_\alpha } +\tr{\hat{\rho}_\MRS (\hat{L}^-_\alpha)^\dag \hat{L}^-_\alpha }.
 \end{aligned}
\end{equation}
For the two-reservoir case considered here, the activity can be split into two parts, $\mathcal{K}^{VV}_\alpha=\mathcal{K}^\mathrm{cross}_\alpha+\mathcal{K}^\mathrm{auto}_\alpha$, where we identify the ``cross'' term as the one that is symmetric under exchange of reservoir indices L$\leftrightarrow$R, and the ``auto'' term as the one that---when multiplied by the relaxation rate $\Lambda$---depends on the two reservoirs separately, 
\begin{subequations}
\begin{align}\label{eq:NESB_Kcross}
 \mathcal{K}^\mathrm{cross}_\alpha= \mathcal{K}^\mathrm{cross}&=\frac{\gamma_\mathrm{L}\gamma_\mathrm{R}}{\Lambda}(n_\mathrm{L}(1+n_\mathrm{R})+n_\mathrm{R}(1+n_\mathrm{L})) ,\\
 \mathcal{K}^\mathrm{auto}_\alpha &=2\frac{\gamma_\alpha^2 }{\Lambda}n_\alpha(1+n_\alpha).
\end{align}
\end{subequations}
Again, the contribution $\mathcal{K}^\mathrm{cross}_\alpha$ plays an important role in the zero-frequency noise of the excitation current~\cite{Landi2024Apr}.
The noise in the excitation and the energy currents are given by
\begin{equation}
\begin{aligned}
 S^{(N)}_\mathrm{L}=\frac{S^{(E)}_\mathrm{L}}{(\hbar\Delta)^2} 
 & = \mathcal{K}^\mathrm{cross}_\alpha - \frac{2 \left(I^{(N)}_\mathrm{L}\right)^2 }{\Lambda}.
\end{aligned}
\end{equation}
The entropy-production rate is given by the Clausius relation in the steady state, Eq.~\eqref{eq:entropy_clausius}, and takes the form
\begin{equation}\label{eq: sigma NESB}
 \sigma= \kB\ln\left[\frac{ n_\mathrm{L} (1+ n_\mathrm{R}) }{ n_\mathrm{R} (1+n_\mathrm{L})} \right] I^{(N)}_\mathrm{L}.
\end{equation}
This entropy production rate turns out to be proportional to the logarithm of the ratio between the two rates constituting $\mathcal{K}^\mathrm{cross}_\alpha$, see Eq.~\eqref{eq:NESB_Kcross}. Furthermore, it should be noted that the dynamics in the weak coupling limit are effectively classical and the steady-state density matrix commutes with the system Hamiltonian $[\hat{\rho}_\MRS,\hat{H}_\MRS]=0$. This means that the system cannot violate the classical, \textit{global} TUR, KUR, or TKUR~\cite{Prech2025Jan,VanVu2025Mar}. 
Here, we instead investigate how the presence of anharmonicity (or, in other words, nonlinearity) in the Hamiltonian impacts the validity of the \textit{local} bounds that were derived for bosonic transport in interacting and noninteracting systems and that were applied to a harmonic-oscillator network in Sec.~\ref{sec:HO} above. 
\begin{figure}[b!]
 \centering
 \includegraphics[width=3.3in]{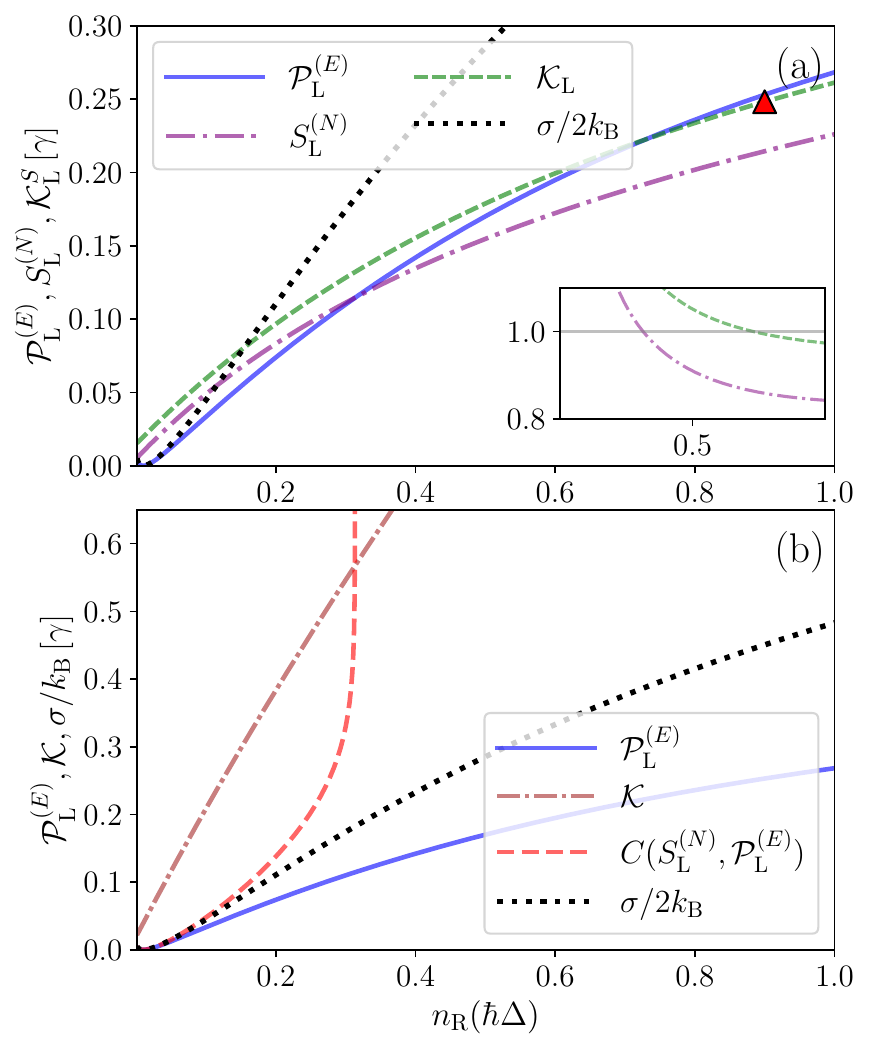}
 \caption{NESB: Energy-current precision $\mathcal{P}^{(E)}_\mathrm{L}$, local activity $\mathcal{K}_\mathrm{L}$, particle current noise $S^{(N)}_\mathrm{L}$, and (for reference) entropy production $\sigma$. Coupling strengths to the reservoirs are symmetric $\gamma_\mathrm{L}=\gamma_\mathrm{R}=\gamma$. Panel (a): right reservoir average occupation $n_\mathrm{R}( \Delta)$ is varied while the average occupation of the left reservoir is kept fixed at $n_\mathrm{L}(\Delta)=0.01$. Panel (b): left reservoir average occupation $n_\mathrm{L}(\Delta)$ is varied while the average occupation of the right reservoir is kept fixed at $n_\mathrm{R}( \Delta)=1$. The insets display violations of local KUR and KUR-like bounds via the ratios $\mathcal{K}_\mathrm{L}/\mathcal{P}^{(E)}_\mathrm{L}$ (dashed green line) and $S^{(N)}_\mathrm{L}/\mathcal{P}^{(E)}_\mathrm{L}$ (dash-dotted purple line). The red triangle indicates a point appearing in Fig.~\ref{fig: boson bound SKUR}.
 }
 \label{fig: bound violation GKSL}
\end{figure}
\subsection{Trade-off relations}
\subsubsection{TUR and local KURs}
We start by showing violations of the local KUR, of the KUR-like bound of Eq.~\eqref{eq:bosonic linear bound} estimating transfer rates using excitation-current noise, as well as of the TKUR in Fig.~\ref{fig: bound violation GKSL} for the energy-current precision. In panel (a), the results are shown as a function of the average occupation of the right reservoir at the energy splitting $\hbar\Delta$, $n_\mathrm{R}(\Delta)$, while keeping the left reservoir at a fixed average occupation $n_\mathrm{L}(\Delta)=0.01$. For the thermal baths considered here, keeping $n_\mathrm{L}( \Delta)=0.01$ and varying $n_\mathrm{R}( \Delta)\in[0,1]$ corresponds to $T_\mathrm{R}/T_\mathrm{L}\in[0,6.66$]. 
 Since all quantities depicted in Fig.~\ref{fig: bound violation GKSL} are linear in the symmetrically chosen coupling $\gamma_\mathrm{L} = \gamma_\mathrm{R} = \gamma$, any observed violation of a bound is independent of the magnitude of~$\gamma$, in the validity regime of the master equation. In panel~(a), it is shown that $\mathcal{P}^{(E)}_\mathrm{L}$ can surpass $\mathcal{K}_\mathrm{L}$ and $S^{(N)}_\mathrm{L}$, meaning that both the local KUR and KUR-like bounds are violated. This becomes particularly clear in the inset, where we plot the ratio between the correlator-based and precision activity and the excitation-current noise and precision. This violation occurs only when the cold contact has a very low occupation number. This is shown in panel (b), where $n_\mathrm{L}$ is modulated at fixed $n_\mathrm{R}(\Delta)=1$. 
 
 When $S^{(N)}_\mathrm{L}<\mathcal{P}^{(E)}_\mathrm{L}=\mathcal{P}^{(N)}_\mathrm{L}$, the transfers obey sub-Poissonian statistics, something which is not possible in linear bosonic transport\footnote{Notice that $S^{(N)}_\mathrm{L}/\mathcal{P}^{(E)}_\mathrm{L}= (S^{(N)}_\mathrm{L}/I^{(N)}_\mathrm{L})^2 $ is the square of the Fano factor due to the connection between particle and energy current.}. 
The sub-Poissonian statistics arise due to anti-bunching of emissions from the TLS, which we show here using the correlation function $g^{(2)}$ for the emissions. The correlation function is calculated as~\cite{Landi2024Apr}
\begin{equation}
 g^{(2)}(\tau) = \frac{\tr{\mathcal{J}^-_\alpha e^{\mathcal{L} \tau} \mathcal{J}^-_\alpha\hat{\rho}_\MRS } }{\tr{\mathcal{J}^-_\alpha \hat{\rho}_\MRS }^2 } = 1-e^{-\Lambda \tau}.
 \end{equation}
Here, $\Lambda$ is the decay rate introduced above and $\mathcal{J}^-_\alpha \bullet= \hat{L}^-_\alpha \bullet (\hat{L}^-_\alpha)^\dag$. Since we here find $g^{(2)}(0)\leq g^{(2)}(\tau)$, the emissions are anti-bunched~\cite{Zou1990Jan,Emary2012Apr,Landi2024Apr}. This is because, as the TLS transmits an excitation to the left bath, it takes some time for the TLS to be re-excited and to transmit a new excitation to the left bath. 
In the weak-coupling limit considered here, we do not observe any cotunnelling, which could otherwise re-induce bunching effects. Importantly, the antibunching observed here is not an effect of quantum coherence; it is simply due to the fact that the TLS system can only host one excitation at a time. Since there is no quantum coherence playing a role in suppressing fluctuations in this regime, we can conclude that the increased precision, compared to bosonic transport in linear systems, is an effect purely due to the anharmonic nature of the NESB. It is thus possible to overcome limits on precision---set for quadratic Hamiltonians in bosonic transport---by utilizing interactions or nonlinearities. There is an interesting subtlety to be pointed out here: while the dynamics of the bosonic transport are fully classical (in the weak coupling limit), the emitted light from the TLS would be in a nonclassical state since $g^{(2)}(0)\leq g^{(2)}(\tau)$~\cite{Kimble1977Sep}.
\begin{figure}[b!]
 \centering
 \includegraphics[width=3.3in]{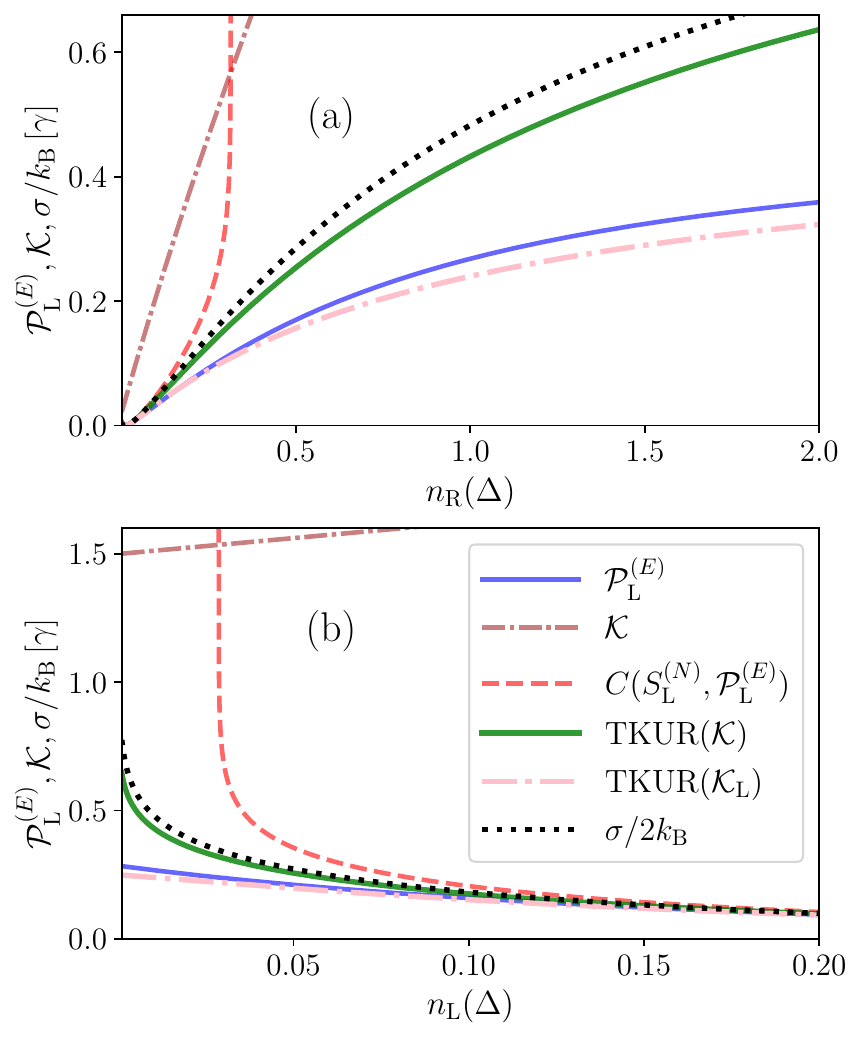}
 \caption{NESB: Energy-current precision $\mathcal{P}^{(E)}_\mathrm{L}$ (blue line), total dynamical activity $\mathcal{K} = \mathcal{K}_\mathrm{L} +\mathcal{K}_\mathrm{R}$ (brown dash-dotted line), entropy-constraint of~\eqref{eq:bosonic linear bound} (red dashed line), TKUR using the total activity (green line), TKUR using local activity (pink dash-dotted line) and total entropy production (black dotted line). Coupling strengths to the reservoirs are symmetric $\gamma_\mathrm{L}=\gamma_\mathrm{R}=\gamma$. Panel (a): right reservoir average occupation $n_\mathrm{R}(\Delta)$ is varied while the average occupation of the left reservoir is kept fixed, $n_\mathrm{L}(\Delta)=0.01$. Panel (b): left reservoir average occupation $n_\mathrm{L}(\Delta)$ is varied while the average occupation of the right reservoir is kept fixed, $n_\mathrm{R}(\Delta)=1$.
 }
 \label{fig: boson bound violation GKSL}
\end{figure}
In Fig.~\ref{fig: boson bound violation GKSL}, we also display the total activity rate $\mathcal{K}$, showing that the global KUR is far from being saturated (while providing a valid bound on $\mathcal{P}^{(E)}_\mathrm{L}$). This is due to the fact that a large part of the total activity is associated with the (hot) contact in which the current is not measured.

\subsubsection{TKUR and entropy-inference bound} 
Furthermore, we show the TKUR using the local and total activities. The standard TKUR using the full activity is valid, being somewhat tighter than both the TUR and KUR, while the TKUR using the local activity is violated in a range of parameter values. 

 In addition, we display the function $C(S^{(N)}_\mathrm{L},\mathcal{P}^{(E)}_\mathrm{L}) $ as a red-dashed line, which for linear or noninteracting bosonic transport provides a \textit{lower} bound on the entropy-production rate $\sigma/2\kB$ via~\eqref{eq:bosonic linear bound}, thereby allowing for inference of the entropy-production rate. We find that the nonlinear nature of the NESB leads to strong violations of~\eqref{eq:bosonic linear bound} with $C(S^{(N)}_\mathrm{L},\mathcal{P}^{(E)}_\mathrm{L}) \not\leq \sigma/2\kB$. In particular, when $S^{(N)}_\mathrm{L}=\mathcal{P}^{(N)}_\mathrm{L}$, the function $C(S^{(N)}_\mathrm{L},\mathcal{P}^{(E)}_\mathrm{L})$ diverges, while the entropy-production rate has a finite value. To see why this happens, we write
 \begin{equation}
 C(S^{(N)}_\MRL, \mathcal{P}^{(E)}_\MRL )=\frac{1}{2} \ln \left[\frac{S^{(N)}_\MRL +|I^{(N)}_\MRL| }{S^{(N)}_\MRL -|I^{(N)}_\MRL| } \right]|I^{(N)}_\MRL|.
 \end{equation}
When $S^{(N)}_\mathrm{L} \leq \mathcal{K}^\mathrm{cross}_\mathrm{L}$, which is the case for the sub-Poissonian noise arising from the nonlinear character of the NESB, $S^{(N)}_\mathrm{L} \pm |I^{(N)}_\mathrm{L}|$ does not provide upper bounds on the rates $ \gamma_\mathrm{L} \gamma_\mathrm{R}n_\mathrm{L}(1+n_\mathrm{R})/\Lambda $ or $\gamma_\mathrm{L} \gamma_\mathrm{R}n_\mathrm{R}(1+n_\mathrm{L})/\Lambda $ which are the rates entering the entropy production~\eqref{eq: sigma NESB}. As a consequence~\eqref{eq:bosonic linear bound} is violated. Instead, in the case of the harmonic network in Sec.~\ref{sec:HO}, we always have $S^{(N)}_\mathrm{L} \geq \mathcal{K}^\mathrm{cross}_\mathrm{L}$, see Eq.~\eqref{eq:part_fluct}, and ~\eqref{eq:bosonic linear bound} is hence a valid lower bound allowing for entropy inference.

However, in the weak-coupling limit considered here, one is, in principle, able to measure the relevant jump rates entering the entropy-production rate, Eq.~\eqref{eq: sigma NESB}, via continuous monitoring, meaning that one can circumvent the need for inferring entropy production from noise and currents, see Ref.~\cite{Menczel2020Sep}.

\subsubsection{Susceptibility-KUR}
Finally, we analyze how the S-KUR of Eq.~\eqref{eq:SKUR}, which is valid for arbitrary quantum systems~\cite{Palmqvist2026Jul} but has been exemplified only for linear systems until now, behaves in the presence of strong nonlinearities. 
Consider the NESB Hamiltonian, where the coupling strength to the left reservoir is controlled by a parameter $\theta$
\begin{equation}\label{eq: ham NESB_theta}
 \hat{H}_\theta =\frac{\hbar \Delta}{2} \hat{\sigma}_z +\sum_{\alpha j} \hbar \omega_{\alpha,j} \hat{b}^\dag_{\alpha,j} \hat{b}_{\alpha,j} + \hat{\sigma}_x (\theta\hat{B}_\MRL +\hat{B}_\MRR).
\end{equation}
This parametrization is inherited by the energy current
\begin{equation}
 \begin{aligned}
 I^{(E)}_\mathrm{L}(\theta)=\frac{ \hbar\Delta\theta^2 \gamma_\mathrm{L}\gamma_\mathrm{R}(n_\mathrm{L}-n_\mathrm{R}) }{\theta^2 \gamma_\MRL (1+2n_\MRL)+ \gamma_\MRR (1+2n_\MRR) }.
 \end{aligned}
\end{equation}
With this we can calculate the susceptibility term $\dot M^{(E)}_\mathrm{L}$ of Eq.~\eqref{eq:response_current}. The standard expression of the energy current is regained setting $\theta=1$. 

\begin{figure}[bt]
 \centering
 \includegraphics[width=3.3in]{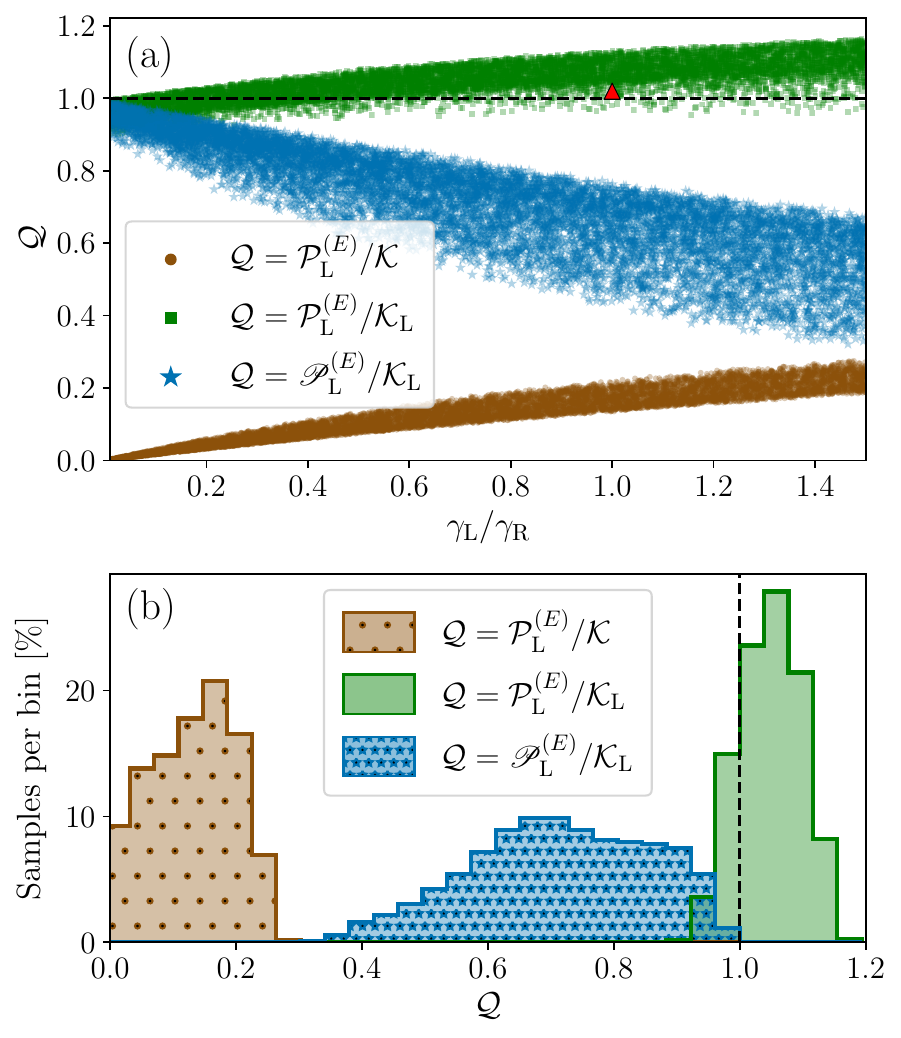}
 \caption{NESB: Violation of local KUR (Eq.~\eqref{eq:local KUR} green squares and histogram) and saturation of S-KUR (\eqref{eq:SKUR} blue stars and histogram) and classical KUR (Eq.~\eqref{eq:global KUR} brown dots and histogram). In panel~(a), we show the ratio between precision and bound as function of the coupling asymmetry for uniformly sampled $n_\MRL(\Delta) \in [10^{-3},10^{-2}]$,
 $n_\MRR(\Delta) \in [0.5,2]$, $\gamma_\MRL/\gamma_\MRR \in [10^{-2},1.5]$ displaying 10000 points. Panel~(b) shows the statistics of values obtained for the ratio between precision and bound. The red triangle indicates a point appearing in Fig.~\ref{fig: bound violation GKSL}.
 }
 \label{fig: boson bound SKUR}
\end{figure}

We apply the S-KUR to the energy transport in the NESB in the weak-coupling Markovian regime. 
In Fig.~\ref{fig: boson bound SKUR}, we compare the violation of the naive local KUR with saturation of the S-KUR and the standard KUR by sampling the coupling asymmetry $\gamma_\MRL /\gamma_\MRR$ and populations $n_\MRL(\Delta)$ and $n_\MRR(\Delta)$ uniformly. Similar to the case of the harmonic oscillator, the standard KUR is far from being saturated, reaching a maximum of $0.247$ while the S-KUR is close to saturation, reaching $0.988$. Unlike the case of the harmonic oscillator, the local KUR breaks, reaching values above $1.162$. Thus, the S-KUR not only offers an extension of the KUR to general quantum transport settings, but it also offers a refinement of the standard KUR in certain classical settings. Moreover, the S-KUR might offer an advantage in experiments where it is only possible or more practical to monitor the activity of one reservoir.

\section{Conclusion}
We have analyzed constraints on the precision of transport observables---in particular, energy transport---in linear and nonlinear systems connected to bosonic reservoirs, exemplified by a harmonic-oscillator network and the nonequilibrium spin-boson model. Constraints are discussed in terms of thermodynamic (TUR) and kinetic (KUR) uncertainty relations, which can be unified in a thermokinetic uncertainty relation (TKUR). For kinetic and thermokinetic uncertainty relations, \textit{local} versions (accounting for activity related to the reservoir of interest only) have previously been introduced, which turn out to be more predictive than the global ones (accounting for the total activity of the transport process).
We have analyzed the predictiveness of these local uncertainty relations for the experimentally relevant models of harmonic oscillators and the NESB in contact with two reservoirs at different temperatures, employing two recently introduced measures of activity---a correlator-based activity~\cite{Blasi2026Jul,Palmqvist2026Jul} and an estimate of activity via the particle-current noise~\cite{Palmqvist2025Jul,Palmqvist2025Oct}. Which of these is more predictive for energy-transport through a harmonic-oscillator system strongly depends on the parameter regimes, in particular on the average reservoir temperatures and the temperature bias. 
By contrast, when introducing nonlinearities (here at the example of the NESB), the local kinetic and thermokinetic uncertainty relations which were developed for linear systems~\cite{Palmqvist2025Jul,Palmqvist2025Oct,Blasi2026Jul} break, when at the same time the temperature bias is large and the temperature of the colder of the two reservoirs is much smaller than the level splitting. This means that anti-bunching, induced by the nonlinearity of the NESB, leads to an increased precision in bosonic energy transport. The recently developed susceptibility-KUR~\cite{Palmqvist2026} instead continues to hold and turns out to be significantly more predictive than the standard classical KUR, in particular for asymmetric coupling strengths.
We also discuss how a recently developed inference bound for entropy production, which is rather tight for the harmonic oscillator network, breaks down in the presence of strong nonlinearities. 

Both the harmonic oscillator network and the NESB are models that characterize well standard setups in circuit quantum electrodynamics (circuit QED). We therefore expect that the insights provided here on precision bounds are of use for the development of future precise devices. Also, the breakdown of the constraints should be detectable in state-of-the-art experiments.

\acknowledgments
We thank Ludovico Tesser for helpful discussions. All the authors acknowledge funding from the DFG - German Research Foundation - under CRC 1277, project-ID 314695032. Furthermore, we acknowledge financial support from the European Research Council (ERC) under the European Union’s Horizon Europe research and innovation program (101088169/NanoRecycle) (D.P.,J.S.), and from the Research Council of Finland through the QTF Centre of Excellence (project No.
336817) (L.M.).

\appendix

\section{Derivation of correlator-based activity for harmonic network}\label{app: deriv gen K}
In order to derive an expression for the correlation based activity we consider a harmonic network partitioned into a central region S and reservoirs labeled by Greek indices $\alpha,\beta,\gamma,\ldots$. The Hamiltonian is
\begin{equation}
 \hat H=\hat H_S + \sum_\beta \left(\hat H_\beta + \hat V_\beta\right),
\end{equation}
with
\begin{equation}
 \hat H_x=\frac{1}{2}\hat p_x^T \hat p_x + \frac{1}{2}\hat u_x^T \mathbf{K}_x \hat u_x,
 \qquad x\in\{\MRS,\alpha,\beta,\ldots\},
\end{equation}
and bilinear couplings
\begin{equation}
 \hat V_\beta = \hat u_\beta^T \mathbf{V}^{\beta \MRS}\hat u_\MRS,
 \qquad
 \mathbf{V}_{\MRS\beta} = (\mathbf{V}_{\beta \MRS})^T.
\end{equation}
The full spring-constant matrix can be written as $\mathbf{K} = \mathbf{k} + \mathbf{V}$ where $\mathbf{k}$ contains the uncoupled central and reservoir blocks, while $\mathbf{V}$ contains only the system-reservoir couplings.

We define the contour-ordered Green's function
\begin{equation}
 \mathbf G(\tau,\tau') = -\frac{i}{\hbar}
 \left\langle T_\mathcal{C}\,\hat u(\tau)\hat u^T(\tau') \right\rangle,
\end{equation}
which obeys the equation of motion~\cite{Wang2014Dec}
\begin{equation}
 \frac{\partial^2 \mathbf G(\tau,\tau')}{\partial \tau^2}=-\mathbf{K}\mathbf G(\tau,\tau')-\delta(\tau,\tau') \mathbf{I},
\end{equation}
and the following Dyson equations
\begin{equation}
\begin{aligned}
 \mathbf G(\tau,\tau') =& \mathbf g(\tau,\tau')+ \int_\mathcal{C} d\tau'' \mathbf g(\tau,\tau'') \mathbf{V}\mathbf G(\tau'',\tau')\\
 =&\mathbf g(\tau,\tau')+ \int_\mathcal{C} d\tau'' \mathbf G(\tau,\tau'')\mathbf{V}\mathbf g(\tau'',\tau')
\end{aligned}
\end{equation}
where $\mathbf g(\tau,\tau')$ are the uncoupled Green's functions. For shorthand, we refer to these convolutions with $ \mathbf G = \mathbf g + \mathbf g\mathbf{V} \mathbf G = \mathbf g + \mathbf G\mathbf{V} \mathbf g$. In the steady state limit where time translation invariance holds, we are able to express the central Green's functions in the frequency domain as~\cite{Wang2014Dec}
\begin{equation}\label{eq:app_GlessGgreat}
 \mathbf G^{</>}_\mathrm{SS}(\omega)=
 \mathbf G^r_\mathrm{SS}(\omega)
 \left(\sum_\beta \mathbf \Sigma^{</>}_\beta(\omega)\right)
 \mathbf G^a_\mathrm{SS}(\omega).
\end{equation}
Moreover, the following identity holds~\cite{Wang2014Dec}
\begin{equation}\label{eq:app_GaGr}
 \mathbf G^a_\mathrm{SS}(\omega)-\mathbf G^r_\mathrm{SS}(\omega)=
 i\sum_\beta \mathbf G^r_\mathrm{SS}(\omega)\mathbf \Gamma_\beta(\omega)\mathbf G^a_\mathrm{SS}(\omega),
\end{equation}
with
\begin{equation}
 \mathbf \Gamma_\beta(\omega)=i\big[\mathbf \Sigma^r_\beta(\omega)-\mathbf \Sigma^a_\beta(\omega)\big].
\end{equation}
Here and in Eq.~\eqref{eq:app_GlessGgreat}, we introduced the self energies defined by $\mathbf \Sigma^\bullet _\alpha=\mathbf{V}_{\MRS \alpha} \mathbf g^\bullet_\alpha \mathbf{V}_{\alpha\MRS }$.
We assume that the retarded and advanced bath self energies are purely imaginary and that the reservoirs are in internal thermal equilibrium
\begin{equation}\label{eq: app self energies}
 \begin{aligned}
 \mathbf \Sigma^r_\alpha(\omega) &=-i \mathbf \Gamma_\alpha(\omega)/2 , & \mathbf \Sigma^a_\alpha(\omega) &=i \mathbf \Gamma_\alpha(\omega)/2 ,\\
 \mathbf \Sigma^<_\alpha(\omega) &=-i n_\alpha(\omega) \mathbf \Gamma_\alpha(\omega) , & \mathbf \Sigma^>_\alpha(\omega) &= -i (1+{n}_\alpha(\omega)) \mathbf \Gamma_\alpha(\omega).
 \end{aligned}
\end{equation}
In addition, we introduce the matrix
\begin{equation}\label{eq:app_def_transmission_matrix}
 \mathbf{T}_{\alpha\beta}(\omega) =
 \mathbf \Gamma_\alpha(\omega)\mathbf G^r_\mathrm{SS}(\omega)\mathbf \Gamma_\beta(\omega)\mathbf G^a_\mathrm{SS}(\omega).
\end{equation}

Next, we calculate the correlator used to define activity~\eqref{eq: corr based K}~\cite{Blasi2026Jul},
\begin{equation}
 \begin{aligned}
 \mathcal{K}^{VV}_\alpha(t)&=\frac{1}{2\hbar^2} \int_{-t}^t dt'\langle\langle \{\hat V_\alpha(t),\hat V_\alpha(t+t') \} \rangle\rangle \\&=\frac{1}{\hbar^2} \Re \int_{-t}^t dt'\langle\langle \hat V_\alpha(t) \hat V_\alpha(t+t') \rangle\rangle.
 \end{aligned}
\end{equation}
We use time-translation invariance, the convolution theorem together with Wick's theorem to write the correlator in the long-time limit $ \mathcal{K}^{VV}_\alpha = \lim_{t\rightarrow\infty} \mathcal{K}^{VV}_\alpha(t)$,
\begin{equation}\label{eq: app ss K trace}
 \begin{aligned}
 \mathcal{K}^{VV}_\alpha
 = -\Re \int^\infty_{-\infty} \frac{d\omega}{2\pi} \Big( &\tr{\mathbf{V}_{\MRS\alpha} \mathbf G^<_{\alpha\alpha}(\omega)\mathbf{V}_{\alpha\MRS} \mathbf G^>_\mathrm{SS}(\omega)} + \\\ +&\tr{ \mathbf{V}_{\alpha\MRS} \mathbf G^<_{\MRS\alpha}(\omega)\mathbf{V}_{\alpha\MRS} \mathbf G^>_{\MRS\alpha}(\omega) } \Big).
 \end{aligned}
\end{equation}
Using the Dyson equation $\mathbf G = \mathbf g + \mathbf G \mathbf{V}\mathbf g$ and Langreth's theorem~\cite{Wang2014Dec}, the steady-state Green's functions can be expressed as
\begin{equation}
 \mathbf G^{</>}_{\MRS\alpha}(\omega) = \mathbf G^r_\mathrm{SS}(\omega) \mathbf{V}_{\MRS\alpha}\mathbf g^{</>}_\alpha(\omega) +\mathbf G^{</>}_\mathrm{SS}(\omega) \mathbf{V}_{\MRS\alpha}\mathbf g^a_\alpha(\omega).
\end{equation}
Moreover, using $\mathbf G = \mathbf g + \mathbf g \mathbf{V} \mathbf G$ we find
\begin{equation}
 \begin{aligned}
 \mathbf G^{</>}_{\alpha\alpha}(\omega) = \mathbf g^{</>}_\alpha (\omega) &+ \mathbf g^{r}_\alpha(\omega) \mathbf{V}_{\alpha \MRS}\mathbf G^{</>}_{\MRS\alpha}(\omega)+ \\ &+\mathbf g^{</>}_\alpha(\omega) \mathbf{V}_{\alpha \MRS}\mathbf G^{a}_{\MRS \alpha}(\omega),
 \end{aligned}
\end{equation}
which together with the Langreth rule $\mathbf G^{r/a}_{\MRS\alpha} (\omega)= \mathbf G^{r/a}_\mathrm{SS}(\omega)\mathbf{V}_{\MRS\alpha}\mathbf g^{r/a}_\alpha(\omega)$ \cite{Wang2014Dec} results in
\begin{equation}
 \begin{aligned}
 \mathbf G^{</>}_{\alpha\alpha}(\omega) = \mathbf g^{</>}_\alpha (\omega) &+ \mathbf g^{r}_\alpha(\omega) \mathbf{V}_{\alpha\MRS}\left[\mathbf G^r_\mathrm{SS}(\omega) \mathbf{V}_{\MRS \alpha}\mathbf g^{</>}_\alpha(\omega)\right]+\\
 &+\mathbf g^{r}_\alpha(\omega) \mathbf{V}_{\alpha \MRS}\left[\mathbf G^{</>}_\mathrm{SS}(\omega) \mathbf{V}_{\MRS \alpha}\mathbf g^a_\alpha(\omega) \right]
 \\ &+\mathbf g^{</>}_\alpha(\omega) \mathbf{V}_{\alpha\MRS} \mathbf G^{a}_\mathrm{SS}(\omega)\mathbf{V}_{\MRS\alpha}\mathbf g^{a}_\alpha(\omega).
 \end{aligned}
\end{equation}

We have now expressed everything in terms of the unperturbed and central Green's functions and what remains is to simplify the expression for $\mathcal{K}^{VV}_\alpha$. Using the cyclicity of the trace and suppressing the frequency arguments, we start with the first trace in Eq.~\eqref{eq: app ss K trace} 
\begin{equation}
 \begin{aligned}
 &\tr{ \mathbf{V}_{\alpha \MRS } \mathbf G^<_{\MRS \alpha} \mathbf{V}_{\alpha \MRS } \mathbf G^>_{\MRS \alpha}}= \\
 &= \tr{ \mathbf G^r_\mathrm{SS} \mathbf \Sigma^<_\alpha \mathbf G^r_\mathrm{SS} \mathbf \Sigma^>_\alpha +\mathbf G^r_\mathrm{SS} \mathbf \Sigma^<_\alpha \mathbf G^>_\mathrm{SS} \mathbf \Sigma^a_\alpha}+ \\
 &\quad+\tr{\mathbf G^<_\mathrm{SS} \mathbf \Sigma^a_\alpha \mathbf G^r_\mathrm{SS} \mathbf \Sigma^>_\alpha+\mathbf G^<_\mathrm{SS} \mathbf \Sigma^a_\alpha \mathbf G^>_\mathrm{SS} \mathbf \Sigma^a_\alpha}.
 \end{aligned}
\end{equation}
The second trace in Eq.~\eqref{eq: app ss K trace} reads as
\begin{equation}\label{eq: app trace 2}
 \begin{aligned}
 &\tr{\mathbf{V}_{\MRS\alpha} \mathbf G^<_{\alpha\alpha}\mathbf{V}_{\alpha \MRS}\mathbf G^>_\mathrm{SS}}\\
 &=\tr{\mathbf \Sigma^<_\alpha \mathbf G^>_\mathrm{SS} + \mathbf \Sigma^r_\alpha \mathbf G^r_\mathrm{SS} \mathbf \Sigma^<_\alpha \mathbf G^>_\mathrm{SS} } +\\ &\quad+\tr{\mathbf \Sigma^r_\alpha \mathbf G^<_\mathrm{SS} \mathbf \Sigma^a_\alpha \mathbf G^>_\mathrm{SS}+\mathbf \Sigma^<_\alpha \mathbf G^a_\mathrm{SS} \mathbf \Sigma^a_\alpha \mathbf G^>_\mathrm{SS}}.
 \end{aligned}
\end{equation}
Moreover, we can write the correlator in an equivalent way
\begin{equation}
 \begin{aligned}
 \mathcal{K}^{VV}_\alpha=
 -\Re \int^\infty_{-\infty} \frac{d\omega}{2\pi}\Big( &\tr{\mathbf{V}_{\MRS\alpha} \mathbf G^>_{\alpha \alpha}(\omega)\mathbf{V}_{\alpha \MRS} \mathbf G^<_\mathrm{SS}(\omega)}+\\+&\tr{ \mathbf{V}_{\MRS \alpha} \mathbf G^>_{\alpha \MRS}(\omega)\mathbf{V}_{\MRS\alpha} \mathbf G^<_{\alpha\MRS}(\omega)}\Big).
 \end{aligned}
\end{equation}
We note that $ \tr{\mathbf{V}_{\MRS \alpha} \mathbf G^>_{\alpha\alpha}\mathbf{V}_{\alpha \MRS} \mathbf G^<_\mathrm{SS}}$ and $ \tr{\mathbf{V}_{\MRS\alpha} \mathbf G^<_{\alpha\alpha}\mathbf{V}_{\alpha \MRS} \mathbf G^>_\mathrm{SS}}$ give the same contribution to $\mathcal{K}^{VV}_\alpha$ after taking the real part and integrating over frequency. This follows from $G^<_{j,k}(t,t’) = G^>_{k,j}(t’,t)$ where $j$ and $k$ refer to indices of the column vectors containing all positions and stationarity.

To continue, the relations Eq.~\eqref{eq:app_GlessGgreat} and Eq.~\eqref{eq:app_GaGr} are used together with Eq.~\eqref{eq: app self energies} to simplify the expressions of the traces. For a detailed account of how each term in the traces is simplified, see Ref.~\cite{Palmqvist2026}. Moreover, to find a symmetric expression for the correlator, the two equivalent expressions are symmetrized
\begin{equation}\label{eq: trace 1 symm}
\begin{aligned}
 &\frac{1}{2}\Big( \tr{\mathbf{V}_{\MRS \alpha} \mathbf G^>_{\alpha\alpha}\mathbf{V}_{\alpha \MRS} \mathbf G^<_\mathrm{SS}} + \tr{\mathbf{V}_{\MRS\alpha} \mathbf G^<_{\alpha\alpha}\mathbf{V}_{\alpha \MRS} \mathbf G^>_\mathrm{SS}}\Big) \\= &-\sum_{\beta} \tr{\mathbf{T}_{\alpha \beta}}\frac{(F_{ \alpha\beta} + F_{\beta \alpha})}{2}- \sum_{\beta \gamma}\tr{\mathbf{T}_{\alpha\beta} \mathbf{T}_{\alpha\gamma}} \frac{F_{\beta \gamma}}{4}+ \\ &+\sum_{\beta \gamma} \tr{\mathbf{T}_{\alpha\gamma} \mathbf{T}_{\alpha\beta}} \frac{(F_{ \alpha\beta} + F_{\beta \alpha})}{4}.
\end{aligned}
\end{equation}
Here we used $F_{ \alpha\beta} (\omega) = n_\alpha(\omega)(1 +n_\beta(\omega))$. Combining this symmetrized expression Eq.~\eqref{eq: trace 1 symm} with Eq.~\eqref{eq: app trace 2} in Eq.~\eqref{eq: app ss K trace} results in
\begin{equation}
 \begin{aligned}
 \mathcal{K}^{VV}_\alpha&=
 \int^\infty_{-\infty} \frac{d\omega}{4\pi} \Big( \mathrm{tr}\Big\{4\mathbf{T}_{\alpha \alpha}(\omega)-\Big(\sum_\beta \mathbf{T}_{\alpha\beta}(\omega)\Big)^2\Big\} F_{\alpha \alpha}(\omega) \\ & \quad\quad\quad\quad+\sum_{\beta\neq\alpha} \tr{\mathbf{T}_{\alpha\beta}(\omega)}(F_{ \alpha\beta} (\omega)+ F_{\beta \alpha}(\omega))\Big)\\
 &=
 \int^\infty_{0} \frac{d\omega}{2\pi}\Big( \mathrm{tr}\Big\{4\mathbf{T}_{\alpha \alpha}(\omega)-\Big(\sum_\beta \mathbf{T}_{\alpha\beta}(\omega)\Big)^2\Big\} F_{\alpha \alpha}(\omega) \\ & \quad\quad\quad\quad+ \sum_{\beta\neq\alpha} \tr{\mathbf{T}_{\alpha\beta}(\omega)}(F_{ \alpha\beta} (\omega)+ F_{\beta \alpha}(\omega))\Big).
 \end{aligned}
\end{equation}

\section{Weak-coupling limit of correlator-based activity}\label{app: weak coupling correlator}

In Ref.~\cite{Blasi2026Jul}, where the correlator-based definition of dynamical activity~\eqref{eq: corr based K} was introduced, it was shown that the definition recovered the standard rate of jumps~\eqref{eq: dynamical activity GKSL} for a quadratic Hamiltonian, in the steady state. Here, we follow a different approach by applying the Born, Markov, and secular approximations to show that Eq.~\eqref{eq: corr based K} recovers the standard GKSL definition also for Hamiltonians that are not quadratic. In addition, our derivation applies outside of the steady state and does not rely upon the assumption that the Fermi or Bose-Einstein functions are energy independent. We consider the Hamiltonian of a system coupled to baths labelled by $\alpha,\beta,\dots$
\begin{equation}
 \hat{H} = \hat{H}_\MRS + \sum_\alpha (\hat{H}_\alpha + \hat{V}_\alpha),
\end{equation}
where the interaction between bath $\alpha$ and the central system is given by
\begin{equation}
 \hat{V}_\alpha = \sum_k \hat{A}_{\alpha k} \otimes \hat{B}_{\alpha k} =\sum_k \hat{A}^\dag_{\alpha k} \otimes \hat{B}_{\alpha k}^\dag.
\end{equation}
Here $\hat{A}_{\alpha k}$ is an operator of the system and $\hat{B}_{\alpha k}$ is an operator of bath $\alpha$. Furthermore, we define $\hat{H}_0 =\hat{H}_\MRS + \sum_\alpha \hat{H}_\alpha $ and $\hat{V}= \sum_\alpha \hat{V}_\alpha$ with the aim of switching to the interaction picture. We define the time evolution operators
\begin{equation}
 \begin{aligned}
 \hat{U}_0(t,0) &= {T}_+\exp\left\{-\frac{i}{\hbar} \int^t_{0} d\tau \hat{H}_0(\tau) \right\} ,\\
 \hat{U}_\mathrm{I}(t,0) &= {T}_+\exp\left\{-\frac{i}{\hbar} \int^t_{0} d\tau \hat{V}^\mathrm{I}(\tau) \right\},
 \end{aligned}
\end{equation}
where $\hat{V}^\mathrm{I}(t)= \hat{U}^\dag_0(t,0)\hat{V}\hat{U}_0(t,0)$. The density operator in the interaction picture at time $t$ is given by $\hat{\rho}^\mathrm{I}(t) = \hat{U}_\mathrm{I}(t,0) \hat{\rho}(0) \hat{U}^\dag_\mathrm{I}(t,0)$. Starting from the Heisenberg picture, we rewrite the correlator in the interaction picture
\begin{equation}
 \begin{aligned}
 &\tr{\hat{V}^\mathrm{H}(t+t') \hat{V}^\mathrm{H}(t) \rho(0) } = \\
 &=\tr{\hat{U}^\dag_\mathrm{I}(t+t',t) \hat{V}^\mathrm{I}(t+t') \hat{U}_\mathrm{I}(t+t',t) \hat{V}^\mathrm{I}(t) \hat{\rho}^\mathrm{I}(t) }.
 \end{aligned}
\end{equation}
We note that the expression for the activity used for GKSL dynamics is second order in the interaction and linear in the density matrix. Since the correlation-based activity~\eqref{eq: corr based K} is second order in $\hat{V}^\mathrm{H}_\alpha(t)$ we expand $\hat{U}^\dag_\mathrm{I}(t+t',t)$ and $\hat{U}_\mathrm{I}(t+t',t)$ to zeroth order, since the rest of the term would only contribute to higher order. Keeping terms which are maximum second order in $\hat{V}^\mathrm{I}$ we get
\begin{equation}
 \begin{aligned}
 &\tr{\hat{V}^\mathrm{H}(t+t') \hat{V}^\mathrm{H}(t) \hat{\rho}(0) } \approx \tr{\hat{V}^\mathrm{I}(t+t') \hat{V}^\mathrm{I}(t) \hat{\rho}^\mathrm{I}(t)}.
 \end{aligned}
\end{equation}
Next, we assume that the spectrum of $\hat{H}_\MRS$ is discrete and denote the eigenvalues of the system Hamiltonian as $\epsilon$ with the projector onto the eigenspace of $\epsilon$ as $\hat{\Pi}(\epsilon)$. We define
\begin{equation}
 \hat{A}_{\alpha k}(\omega) =\sum_{\epsilon'-\epsilon=\hbar\omega} \hat{\Pi}(\epsilon) \hat{A}_{\alpha k} \hat{\Pi}(\epsilon'),
\end{equation}
with the properties~\cite{Breuer2007}
\begin{equation}
 [\hat{H}_\MRS, \hat{A}_{\alpha k}(\omega)]= -\hbar\omega \hat{A}_{\alpha k}(\omega),\quad \hat{A}^\dag_{\alpha k}(\omega) = \hat{A}_{\alpha k}(-\omega). 
\end{equation}
Under the assumption of there being no explicit time dependence, the operators in the interaction picture evolve as
\begin{align}
 \hat{A}^\mathrm{I}_{\alpha k}(\omega;t)&=e^{i \hat{H}_\MRS t/\hbar } \hat{A}_{\alpha k}(\omega) e^{-i \hat{H}_\MRS t/\hbar } = e^{-i \omega t}\hat{A}_{\alpha k}(\omega),\\
 \hat{B}^\mathrm{I}_{\alpha k}(t)&=e^{i \hat{H}_\alpha t/\hbar } \hat{B}_{\alpha k} e^{-i \hat{H}_\alpha t/\hbar }.
\end{align}
Moreover, we assume $\langle \hat{B}^\mathrm{I}_\alpha(t)\rangle=0$ and weak coupling, such that we can use the Born approximation $\hat{\rho}^\mathrm{I}(t) \approx \hat{\rho}^\mathrm{I}_\MRS(t)\bigotimes_\alpha \hat{\rho}_\alpha$, where $\hat{\rho}_\alpha$ are density matrices of the baths. This means that the correlator of the full interaction Hamiltonian decomposes into a sum
\begin{equation}
 \tr{\hat{V}^\mathrm{I}(t+t') \hat{V}^\mathrm{I}(t) \hat{\rho}^\mathrm{I}(t)} = \sum_\alpha \tr{\hat{V}^\mathrm{I}_\alpha(t+t') \hat{V}^\mathrm{I}_\alpha(t) \hat{\rho}^\mathrm{I}(t)}.
\end{equation}
Thus, we focus on the correlator-based activity~\eqref{eq: corr based K} for a single reservoir and approximate
\begin{equation}
 \begin{aligned}
 \mathcal{K}^{VV}_\alpha (t)&\approx \Re \sum_{kk'} \int^t_{-t}\frac{dt'}{\hbar^2} \Big( \tr{\hat{A}^\mathrm{I}_{\alpha k'} (t+t') \hat{A}^\mathrm{I}_{\alpha k} (t) \hat{\rho}^\mathrm{I}_\MRS(t)} \times \\ &\quad\quad\quad\quad\quad\quad\times \tr{ \hat{B}^{\mathrm{I}}_{\alpha k'} (t+t') \hat{B}^{\mathrm{I}}_{\alpha k} (t) \hat{\rho}_\alpha} \Big)\\
 &=\Re \sum_{kk'} \sum_{\omega\omega'} e^{-i(\omega +\omega')t} \tr{\hat{A}_{\alpha k'}(\omega') \hat{A}_{\alpha k}(\omega) \hat{\rho}^\mathrm{I}_\MRS(t)} \times\\
 &\quad\quad\quad\times \int^t_{-t} \frac{dt'}{\hbar^2}e^{-i \omega' t'} \tr{\hat{B}^{\mathrm{I} \dag}_{\alpha k'} (t+t') \hat{B}^{\mathrm{I}}_{\alpha k} (t) \hat{\rho}_\alpha}.
 \end{aligned}
\end{equation}

Next, we use the Markov approximation where the bath correlation functions decay fast compared to the time scale of the system dynamics and extend the integration limits to $\pm\infty$. We introduce the bath correlation functions
\begin{equation}
\begin{aligned}
 \gamma^\alpha_{k'k}(\omega) &= \int^\infty_{-\infty} dt'e^{i \omega t'} \tr{\hat{B}^\mathrm{I}_{\alpha k'} (t') \hat{B}^\mathrm{I}_{\alpha k} (0) \hat{\rho}_\alpha} ,
\end{aligned}
\end{equation}
and employ the secular approximation, neglecting any oscillating terms where $\omega +\omega' \neq 0$, yielding
\begin{equation}\label{eq: KVV nondiag}
 \begin{aligned}
 \mathcal{K}^{VV}_\alpha 
 &=\Re \sum_{kk'}\sum_{\omega} \tr{\hat{A}^\dag_{\alpha k'}(\omega) \hat{A}_{\alpha k}(\omega) \hat{\rho}^\mathrm{I}_\MRS(t)} \gamma^\alpha_{k'k}(\omega).
 \end{aligned}
\end{equation}
Note that to get the total number of interactions observed $\mathcal{A}_\alpha(t_\mathrm{M})$ between time $0$ and $t$, one integrates $\mathcal{A}_\alpha(t) = \int_0^{t}d\tau \mathcal{K}^{VV}_\alpha(\tau)$, meaning that the rotating terms do not contribute.

The matrix $\gamma^\alpha_{k'k}(\omega)$ is hermitian and positive, following from Bochner's theorem~\cite{Breuer2007} and is diagonalized by
\begin{equation}
 \gamma^\alpha_{k'k}(\omega) = \sum_j u^\alpha_{k'j}(\omega) \gamma^\alpha_j(\omega) (u^{\alpha}_{kj}(\omega))^*,
\end{equation}
with $\gamma^\alpha_j(\omega) \geq 0$ being the $j$-th eigenvalue of $\gamma^\alpha(\omega)$ with the corresponding eigenvector $\boldsymbol{u}_j(\omega)$. Defining the usual jump operators
\begin{equation}
 \hat{L}^\dag_{\alpha j}(\omega) = \frac{\sqrt{ \gamma^{\alpha}_j(\omega) }}{\hbar} \sum_{k} u^\alpha_{j k}(\omega) A^\dag_{\alpha k}(\omega),
\end{equation}
Eq.~\eqref{eq: KVV nondiag} is expressed as
\begin{equation}\label{eq: KVV diag}
 \begin{aligned}
 \mathcal{K}^{VV}_\alpha(t) = \sum_\omega\sum_j \tr{\hat{L}^\dag_{\alpha j}(\omega) \hat{L}_{\alpha j}(\omega) \hat{\rho}^\mathrm{I}_\MRS(t)}.
 \end{aligned}
\end{equation}
Eq.~\eqref{eq: KVV diag} is the standard rate of jumps induced by $\alpha$ in the context of GKSL dynamics. The total number of jumps is obtained from 
\begin{equation}
 \AGKSL(t) = \int^t_0d\tau \sum_\alpha \mathcal{K}_\alpha^{VV}(\tau).
\end{equation}
This derivation shows how the correlator-based activity~\eqref{eq: corr based K} reduces to the usual jump definition in GKSL dynamics.

\bibliography{refs.bib}
\end{document}